\documentstyle[epsfig, subfigure, lscape]{mn}

\def\ltapprox{\raise 2pt \hbox {$<$} \kern-1.1em \lower 5pt \hbox {$\approx$}}
\def\ltsim{\raise 2pt \hbox {$<$} \kern-1.1em \lower 4pt \hbox {$\sim$}}
\def\gtsim{\raise 2pt \hbox {$>$} \kern-1.1em \lower 4pt \hbox {$\sim$}}

\title{Alfv{\'e}nic reacceleration of relativistic particles 
in galaxy clusters: MHD waves, leptons and hadrons}
\author[G. Brunetti et al.]
      {G. Brunetti,$^1$
       P. Blasi,$^{2,3}$
       R. Cassano,$^{4,1}$ 
       S. Gabici$^{5}$\\
       $^1$ Istituto di Radioastronomia del CNR, via Gobetti 101,
       I--40129 Bologna, Italy \\
       $^2$ INAF/Osservatorio Astrofisico di Arcetri,
       Largo E. Fermi 5, I-50125 Firenze, Italy \\
       $^3$ INFN/Sezione di Firenze \\
       $^4$ Dipartimento di Astronomia,Universita' di
       Bologna, via Ranzani 1, I-40127 Bologna, Italy\\
       $^5$ Dipartimento di Astronomia e Scienza dello Spazio, 
       Universita' di Firenze, Largo E. Fermi 5, I-50125 Firenze, 
       Italy\\
}

\begin{document}
\maketitle

\begin{abstract}

There is a growing evidence that extended radio halos are most likely 
generated by electrons reaccelerated via some kind of turbulence generated 
in the cluster volume during major mergers. 
It is well known that Alfv\'en waves channel most of their energy flux 
in the acceleration of relativistic particles. Much work has been done 
recently to study this phenomenon and its consequences for the 
explanation of the observed non-thermal phenomena in clusters of galaxies.
We investigate here the problem of particle-wave interactions in the most
general situation in which relativistic electrons, thermal protons
and relativistic protons exist within the cluster volume. The interaction
of all these components with the waves, as well as the turbulent cascading
and damping processes of Alfv\'en waves, are treated in a fully time-dependent
way. This allows us to calculate the spectra of electrons, protons and waves 
at any fixed time. The {\it Lighthill} mechanism is 
invoked to couple the fluid turbulence, supposedly injected during 
cluster mergers, to MHD turbulence. 
We find that present observations
of non-thermal radiation from clusters of galaxies are well described
within this approach, provided the fraction of relativistic hadrons in 
the intracluster medium (ICM) is smaller than $5-10\%$.
\end{abstract}

\begin{keywords}
acceleration of particles - radiation mechanisms: non--thermal -
galaxies: clusters: general -
radio continuum: general - X--rays: general
\end{keywords}

\maketitle

\section{Introduction}

There is now firm evidence that the ICM is a mixture of hot gas, magnetic 
fields and relativistic particles. While the hot gas results in thermal
bremsstrahlung X-ray emission, relativistic electrons
generate non-thermal radio and hard X-ray radiation. The amount of the
energy budget of the intracluster medium in the form of high energy 
hadrons can be large, due to the phenomenon of confinement of cosmic rays 
over cosmological time scales 
(V\"{o}lk et al. 1996;
Berezinsky, Blasi \& Ptuskin 1997). 
Nevertheless, the gamma radiation that would allow us to 
infer the fraction of relativistic hadrons in clusters has not been detected 
as yet (Reimer et al., 2003).
 
The most important evidence for relativistic electrons in clusters
of galaxies comes from the diffuse synchrotron radio emission observed in 
about $35\%$ of the clusters selected with X--ray luminosity $> 10^{45}$ 
erg s$^{-1}$ (e.g., Feretti, 2003). The diffuse emission comes in two 
flavors, referred to as radio halos (and/or radio mini--halos) when the 
emission appears concentrated at the center of the cluster, and radio relics 
when the emission comes from the peripherical regions of the cluster.

The difficulty in explaining the extended radio halos arises from the 
combination of their $\sim$Mpc size, and the relatively short radiative 
lifetime of the radio emitting electrons. Indeed, the diffusion time 
necessary for the radio electrons to cover such distances is orders of 
magnitude larger than their radiative lifetime.
As proposed first by Jaffe (1977), a solution to this puzzle would be
provided by continuous {\it in situ} reacceleration of the relativistic 
electrons on their way out. This possibility was studied more quantitatively
by Schlickeiser et al. (1987) who successfully reproduced the integrated 
radio spectrum of the radio halo in the Coma cluster. In the framework of 
the {\it in situ} reacceleration model, Harris et al. (1980) first suggested 
that cluster mergers might provide the energetics necessary to reaccelerate 
the relativistic particles.

An alternative to the reacceleration scenario was put forward by 
Dennison (1980), who suggested that relativistic electrons may be
produced {\it in situ} by inelastic proton-proton collisions through 
production and decay of charged pions. This model is known in the
literature as the {\it secondary electron model}.

Diffuse radio emission is not the only evidence of non-thermal 
activity in the ICM. Additional evidence, although limited to a few
cases, comes from the detection of extreme ultra-violet (EUV) excess 
emission (e.g., Bowyer et al. 1996; Lieu et al., 1996; Bergh\"ofer et 
al., 2000; Bonamente et al., 2001), and of hard X-ray (HXR) excess 
emission in the case of the Coma cluster, A2256 and possibly A754 
(Fusco--Femiano et al., 1999, 2000, 2003, 2004; 
Rephaeli et al., 1999; Rephaeli 
\& Gruber, 2002, 2003)\footnote{The evidence for HXR excess in the 
case of the Coma
cluster has been recently disputed by Rossetti \& Molendi (2004)}. 
While, with the exception of the Coma and Virgo clusters, the detection 
of EUV emission is still controversial (e.g., Bergh\"ofer et al., 2000), 
the detection of HXRs appears robust as it is claimed independently 
by different groups and with different X--ray observatories
(BeppoSAX, RXTE). If these excesses are indeed of non-thermal origin,
they may be explained in terms of IC scattering of relativistic
electrons off the photons of the cosmic microwave background (CMB)
(Fusco--Femiano et al., 1999, 2000; Rephaeli et al., 1999; V\"olk \& 
Atoyan 1999; Blasi 2001; Brunetti et al.2001a; 
Fujita \& Sarazin 2001; Petrosian 2001; Kuo et al., 2003).
Alternatively HXRs might also result from bremsstrahlung emission of 
electrons in the thermal gas whose spectrum is not a perfect Maxwellian,
but rather has a supra-thermal tail (e.g., Ensslin, Lieu, Biermann 1999;
Blasi 2000; Dogiel 2000; Sarazin \& Kempner 2000). Blasi (2000) calculated
the spectrum of the electrons as modified by the resonant interaction 
with MHD waves by solving a time-dependent Fokker-Planck equation and reached
the conclusion that the intracluster medium is heated and at the same time
develops a supra-thermal tail due to the resonant particle-wave interaction.
An energy injection comparable with the whole luminosity of a merger, lasting
for about a billion years was required in this calculation (Blasi 2000)
in order to explain observations. A shorter allowed duration (about $10^8$ 
years) of the event was estimated by Petrosian (2001). 

Both the IC model and the bremsstrahlung interpretation have problems:
the first one would require cluster magnetic field strengths smaller than 
those inferred from measurements of the Faraday rotations (e.g., Carilli 
\& Taylor 2002). The second one would require a large amount of
energy to maintain a substantial fraction of the thermal electrons far 
from the thermal equilibrium for more than a few $10^8$yrs (Blasi, 2000; 
Petrosian, 2001).

The origin of radio halos is still an open problem. In principle, it is 
known that the fine radio properties of the radio halos may be naturally 
accounted for in models in which electrons are directly accelerated (primary
electrons) whereas they are difficult to reproduce in models in which the
electrons are secondary products of hadronic interactions (Brunetti, 2002). 
On the other hand, the details of the physics used in the primary electron 
models are relatively uncertain.
In principle, merger shocks can accelerate relativistic electrons 
to produce large scale synchrotron radio emission (e.g., Roettiger 
et al., 1999; Sarazin 1999; Takizawa \& Naito, 2000), however, the 
radiative life--time of the emitting electrons diffusing away from these
shocks is so short that they would just be able to produce
relics and not Mpc scale radio halos (e.g., Miniati et al, 2001).
In addition, a number of papers (Gabici \& Blasi 2003; Berrington \& 
Dermer 2003) have recently pointed out that the Mach 
number of the typical shocks produced during major merger events is 
too low to generate non--thermal radiation with the observed fluxes
and spectra.

Re--acceleration of a population of relic electrons by turbulence powered 
by major mergers is suitable to explain the very large scale of 
the observed radio emission and is also a promising possibility 
to account for the fine radio structure of the diffuse emission
(Brunetti et al., 2001a,b).
There are a number of possibilities to channel the energy of the
turbulence in the acceleration of fast particles, namely
via Magneto-Sonic (MS) waves, via magnetic Landau damping (e.g., 
Kulsrud \& Ferrari 1971), via Lower Hybrid (LH) waves (e.g., Eilek 
\& Weatherall 1999) or via Alfv\'en waves.
Since Alfv\'en waves are likely to be able to transfer most of their
energy into relativistic particles, they have received much attention 
in the last few years. In this framework for instance Ohno, Takizawa 
and Shibata (2002) developed a time-independent model for the acceleration
of the relativistic electrons expected in radio halos through magnetic 
turbulence. The authors studied the acceleration of continuously injected
relativistic electrons by Alfv\'en waves with a power law spectrum and 
applied this model to the case of the radio halo in the Coma cluster. 
More recently, Fujita, Takizawa and Sarazin (2003) studied the effect of 
Alfv\'enic acceleration of relativistic electrons in clusters of galaxies.
These authors invoked the {\it Lighthill} theory to establish a connection
between the large scale fluid turbulence and the radiated MHD waves.
The electron and MHD-wave spectra adopted by Fujita et al.(2003) are 
obtained via a self-similar approach by requiring that the spectra are 
described by two power laws.

These approaches have two intrinsic limitations: the first one is in 
the assumption, mentioned above, that all spectra are time-independent and 
that the turbulence spectrum is a power law.
The second is that they neglect, as all other previous approaches did,
the effect of relativistic hadrons in the ICM: it is well known that 
the interaction of the Alfv\'en waves with relativistic particles is, in 
general, more effective for protons than for electrons (e.g., Eilek 1979).
It is also well known that the presence of a significant energy budget 
in the form of relativistic particles can significantly affect the spectrum 
of the Alfv\'en waves through damping. In fact, this damping occurs even 
on the thermal protons in the ICM, another effect which was never
included in previous calculations.

The calculations presented here provide a self-consistent time-dependent 
treatment of the non-linear coupling of Alfv\'en waves, relativistic 
electrons, thermal and relativistic protons. The results 
previously appeared in the literature can be obtained as special 
cases of our very general approach, which is in principle applicable
to scenarios other than clusters of galaxies.

The paper is organized as follows: in Sect.~2 we discuss the energy 
losses of relativistic particles and the presence of these particles 
in galaxy clusters. In Sect.~3 we discuss the physics of Alfv\'en waves, 
their generation in galaxy clusters and their interaction with particles.
In Sect.~4 we discuss the physics of the coupling between Alfv\'en waves, 
leptons and hadrons and derive the time evolution of waves and 
particles as a function of the physical conditions typical of the ICM. 
In Sect.~5 we apply our general formalism to calculate the fluxes of
radio radiation and hard X--ray tails in galaxy clusters.

Throughout the paper we adopt $H_0 = 50$ km s$^{-1}$ Mpc$^{-1}$;
if not specified all the quantities are given in c.g.s. units.

\section{Basic equations and assumptions}

\subsection{Energy losses of relativistic particles 
in the ICM}

In this Section we give a short summary of the main energy loss 
channels that may be important for electrons and protons in the ICM.

\subsubsection{Electrons}

Relativistic electrons with momentum $p = m_{\rm e} c \gamma$
in the ICM lose energy through ionization losses and Coulomb 
collisions (Sarazin 1999):
\begin{equation}
\left( {{ d p }\over{d t}}\right)_{\rm i} 
=- 3.3 \times 10^{-29} n_{\rm th}
\left[1+ {{ {\rm ln}(\gamma/{n_{\rm th}} ) }\over{
75 }}\right] 
\label{ion}
\end{equation}
\noindent
where $n_{\rm th}$ is the number density of the thermal plasma.
Relativistic electrons lose energy via synchrotron emission and inverse 
Compton scattering (ICS):
\begin{equation}
\left( {{ d p }\over{d t}}\right)_{\rm rad}
=- 4.8 \times 10^{-4} p^2
\left[ \left( {{ B_{\mu G} }\over{
3.2}} \right)^2 {{ \sin^2\theta}\over{2/3}}
+ (1+z)^4 \right] 
\label{syn+ic}
\end{equation}
\noindent
where $B_{\mu G}$ is the magnetic field strength in $\mu G$ and $\theta$ 
is the pitch angle of the emitting electrons; in case of efficient 
isotropization of the electron momenta it is possible to average over
all possible pitch angles, so that  $<\sin^2\theta> = 2/3$.
It is well known that in the typical conditions of the ICM radiation losses 
are the most important for electrons with Lorentz factor $\gamma\gg 100$ 
while Coulomb losses dominate at lower energies (Sarazin 1999,2001; Brunetti 
2002). The lifetime of relativistic electrons, defined as $\tau \sim \gamma/ 
\dot{\gamma}$, can be easily estimated from Eqs.(\ref{ion}--\ref{syn+ic}) as:
\begin{eqnarray}
\tau_{\rm e}({\rm Gyr}) \sim
4 \times \Big\{
{1\over 3}
\Big( {{\gamma}\over{300}} \Big)
\left[ \left( {{ B_{\mu G} }\over{
3.2}} \right)^2 {{ \sin^2\theta}\over{2/3}}
+ (1+z)^4 \right]
\nonumber\\
+
\Big( {{ n_{\rm th} }\over{
10^{-3} }} \Big)
\Big({{ \gamma }\over{
300 }} \Big)^{-1}
\left[1.2 + {1\over{75}}
\ln \Big( {{\gamma/300}\over{
n_{\rm th}/10^{-3} }} \Big) \right]
\Big\}^{-1}.
\label{tau1/2}
\end{eqnarray}
\noindent

\subsubsection{Protons}

The main channel of energy losses for relativistic protons is represented 
by inelastic proton-proton collisions. The timescale associated with this 
process is :
\begin{equation}
\tau_{pp} = 
{ 1\over
{{ 
n_{\rm th} \sigma_{\rm pp} c
}} }
\sim 10^{18} \Big(
{{ n_{\rm th} }\over{ 10^{-3}}}
\Big)^{-1} ~.
\label{taupp}
\end{equation}
\noindent
Inelastic ${\rm pp}$ scattering is weak enough to allow for the accumulation 
of protons over cosmological times (Berezinsky, Blasi \& Ptuskin 1997). 
The rare interactions with the ICM may generate an appreciable flux of 
gamma rays and neutrinos, in addition to a population of secondary 
electrons (Blasi \& Colafrancesco 1999). The process of pion production 
in $pp$ scattering is a threshold reaction that requires protons with 
kinetic energy larger than $\sim 300$ MeV. 

Protons which are more energetic than the thermal electrons, namely 
protons with velocity $\beta > \beta_c=(3/2 m_e/m_p)^{1/2} \beta_e$ 
($\beta_e$ here is the velocity of the thermal electrons, $\beta_e 
\simeq 0.18 (T/10^8 K)^{1/2}$) lose energy due to Coulomb interactions.
If we define $x_m = \left( {{ 3 \sqrt{\pi}}\over{4}} \right)^{1/3}\beta_e$, 
we can write (e.g., Schlickeiser, 2002):
\begin{equation}
{{ d p }\over{dt}} \simeq
- 1.7 \times 10^{-29}
\left( {{n_{\rm th}}\over{10^{-3}}} \right)
{{
\beta }\over{
x_m^3 + \beta^3 }}
\label{coulomb_p}
\end{equation}
\noindent
with the following asymptotic behaviour:
\begin{equation}
{{ d p }\over{dt}} 
\propto
\left( {{n_{\rm th}}\over{10^{-3}}} \right)
\times 
\left\lbrace \begin{array}{lll}
\textrm{Const.} & \textrm{for} & p>> m c \\
p^{-2} & \textrm{for} & mc x_m < p << m c \\
p      & \textrm{for} & mc \beta_c < p < mc x_m
\end{array}
\right.
\label{propc}
\end{equation}

\noindent
The timescale associated with Coulomb collisions (in the case $mc x_m < p \ll 
m c$) can be therefore written as:
\begin{equation}
\tau_{C}
\sim 2.5 \times \tau_{pp} \Big(
{{ p }\over{ m_{\rm p} c }}
\Big)^3.
\label{timescaleppi}
\end{equation}
\noindent
For trans-relativistic and sub-relativistic protons  
this channel can easily become
the main channel of energy losses in the ICM.

\subsection{Origin and spectrum of the relic relativistic particles}

In this section we briefly discuss the mechanisms responsible for 
the injection of cosmic rays in galaxy clusters, more extended discussions
have been previously presented by Berezinsky, Blasi \& Ptuskin (1997), 
Kronberg (2002), Biermann et al. (2002) and Jones et al. (2002).

Collisionless shocks are generally recognized as efficient particle
accelerators through the so-called ``diffusive shock acceleration'' (DSA) 
process (Drury, 1983; Blandford \& Eichler 1987).
This mechanism has been invoked several times as the main acceleration
process in clusters of galaxies that have been involved in a merger
event (Takizawa \& Naito, 2000; Blasi 2001;
Miniati et al., 2001; 
Fujita \& Sarazin 2001). For the case of standard newtonian shocks, 
relevant for clusters of galaxies, the spectrum of the accelerated particles 
can be shown to be a power-law with a slope that is independent of the 
details of the diffusion in the shock vicinity, and depends only on the 
compression factor $r$ at the shock, $s=(r+2)/(r-1)$.

The injection of CR ions at shocks is generally computed in the {\it test 
particle limit} while the injection efficiency is sometimes just assumed
as a free parameter, while in other cases it is estimated according to 
the so-called  {\it thermal leakage} model (e.g., Kang \& Jones, 1995).

There is still some debate on the typical Mach number of the shocks developed 
in the ICM during cluster mergers. Some results from numerical simulations 
suggest the presence of a large fraction of high Mach number shocks in 
cluster mergers (Miniati et al., 2000, 2001). Semi--analytical 
calculations (Gabici \& Blasi 2003; Berrington \& Dermer 2003) have pointed 
out that the Mach numbers of the shocks related to major mergers are expected
to be of order unity. Recent numerical simulations (Ryu et al., 2003) seem 
to find more weak shocks than in Miniati et al.(2000, 2001). The comparison
however with analytical calculations appears difficult because of a different
classification of the shocks in the two approaches. Moreover the numerical 
simulations of Ryu et al. (2003) also find a class of high Mach number
shocks that are related to gravitationally unbound structures, not included
in semi-analytical calculations. On the other hand, the weakness of the 
merger-related shocks seems to be also suggested by the few observations 
in which the Mach number of the shock can be measured (e.g., Markevitch 
et al., 2003). If shocks related to major mergers are indeed weak, the 
spectra of the accelerated particles are typically too steep to be relevant 
for nonthermal phenomena in clusters of galaxies.

A relevant contribution to the injection of cosmic rays in clusters of 
galaxies may come from Active Galactic Nuclei (AGN). AGNs indeed inject in 
the ICM a considerable amount of energy in relativistic particles and also 
in magnetic fields, likely extracted from the accretion power of their central 
black hole (Ensslin et al., 1997). It should be stressed that the presence
of relativistic plasma in AGNs (radio lobes and jets) is {\it directly} 
observed because of the synchrotron and inverse Compton emission of accelerated
electrons. 

Finally, powerful Galactic Winds (GW) can inject relativistic particles and 
magnetic fields in the ICM (V\"olk \& Atoyan 1999). Although the present day 
level of starburst activity is low, it is expected that these winds were more 
powerful during starburst activity in early galaxies. Some evidence that 
powerful GW were more frequent in the past comes from the observed iron 
abundance in galaxy clusters (V\"olk et al. 1996).

\begin{figure}
\resizebox{\hsize}{!}{\includegraphics{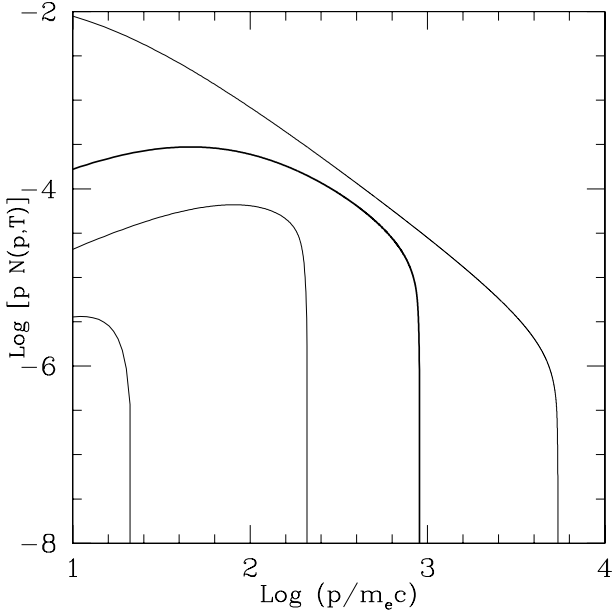}}
\caption[]{Electron spectrum at $z=0$ injected as a single
burst at $z_i=$0.01, 0.1, 0.3, 0.5 (from right to left) adopting
injection spectrum $Q(p) \propto p^{-2.5}$ and maximum Lorentz 
factor $\gamma_{max} = 10^{4}$.
The calculations are carried out for $n_{th}=10^{-3}$cm$^{-3}$ and 
$B=1 \mu$G.}
\end{figure}

Since most of the scenarios discussed above imply power law spectra of 
relativistic particles at the injection sites, in the following we will
restrict our calculation to this case. It is worth recalling that transport 
effects and energy losses modify the shape of these spectra, that are 
not expected to be power laws at later times.

\subsubsection{Electron Spectrum}

In the conditions typical of the ICM, ultra-relativistic electrons rapidly
cool down through ICS and synchrotron emission, and accumulate in the region
of Lorentz factors $\gamma \sim 100-500$ where they may stay for a few 
billion years before cooling further down in energy through Coulomb 
scattering and eventually thermalize.

The kinetic equation that describes these losses and the continuous injection
of electrons is (Kardashev, 1962):
\begin{equation}
{{\partial N_{\rm e}(p,t)}\over{\partial t}}=
{{\partial }\over{\partial p}}
\Big(
(
{ {{ d p }\over{dt}} }_{rad} + { {{ d p }\over{dt}} }_{i}
)
N_{\rm e}(p,t) \Big) +Q_{\rm e}(p)
\label{continuitae}
\end{equation}
\noindent
where ${ d p/ dt }_{\rm i}$, and ${ d p/ dt }_{\rm rad}$ 
are given by Eqs.(\ref{ion})
and (\ref{syn+ic}) respectively and $Q_{\rm e}(p)$ represents the injection
term. The time evolution of relativistic electrons in the ICM during 
cosmological times has been investigated in some detail by Sarazin (1999) by
making use of the numerical solutions of Eq. (\ref{continuitae}). In Fig. 1 we 
plot the spectrum of relativistic electrons measured at the present time and
injected in a single event at $z=z_i$ in the cluster volume.
The calculations are shown for different values of $z_i$: the spectra flatten 
with increasing $z_i$ (typical conditions of the ICM are adopted, as described
in the figure caption). Most electrons injected at $z_i > 0.2$ get thermalized.

In the case of continuous time-independent injection, it is well known that
an equilibrium spectrum of the electrons is achieved, that can be written as
follows:
\begin{equation}
N_{\rm e}(p,z) \propto
\cases{
p^{-s({\cal M})+1},& if $p<<p_{\rm eq}(z)$; \cr
p^{-s({\cal M})-1} & if $p>>p_{\rm eq}(z)$. \cr}
\label{asintotie}
\end{equation}

\noindent
where $p_{\rm eq}$ is close to the momentum at which the Coulomb losses 
dominate over the radiative losses, namely:
\begin{equation}
p_{\rm eq}(z) \sim
300 \times m_{\rm e} c \Big( 
{{ n_{\rm th} }\over{
10^{-3}
}} \Big)^{1/2}
(1+z)^{-2}
\label{peqz}
\end{equation}

More realistic injection histories of relativistic electrons can be
easily implemented in this kind of calculation.

\subsubsection{Protons}

Both timescales of energy losses and diffusion out of the cluster volume
are larger than the Hubble time for most of the cosmic ray protons in the ICM
(V\"olk et al. 1996; Berezinsky, Blasi \& Ptuskin 1997), although the 
energy at which confinement becomes inefficient is rather sensitive on 
the adopted diffusion model.

As stressed above, mildly and sub- relativistic protons may be significantly 
affected by Coulomb energy losses, which in turn change the particle spectrum 
with respect to the injection spectrum.

\begin{figure}
\resizebox{\hsize}{!}{\includegraphics{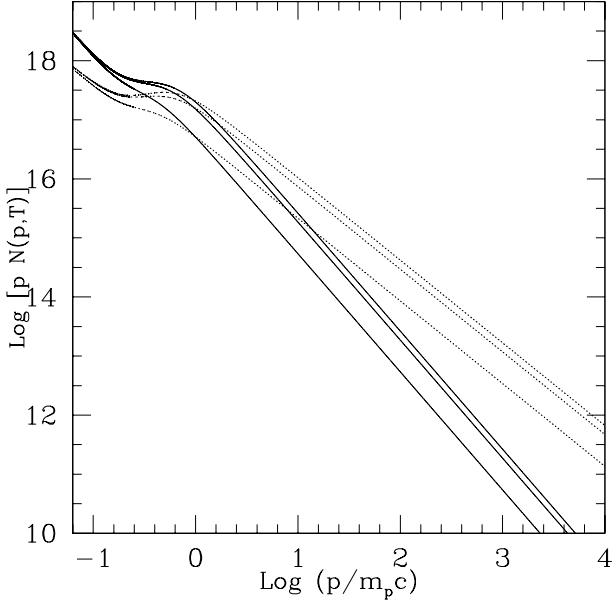}}
\caption[]{Present-epoch spectrum of the cosmic ray protons continuously 
injected (with $p > 0.1\, m_p c$) in the ICM starting from $z_i$.
The spectra are plotted for $s=$2.4 (dotted lines) and 
3.0 (solid lines) and for $z_i=$0.1, 0.5, 1 (from bottom to top). 
Calculations are carried out assuming $n_{th}=10^{-3}$cm$^{-3}$.}
\end{figure}

\begin{figure}
\resizebox{\hsize}{!}{\includegraphics{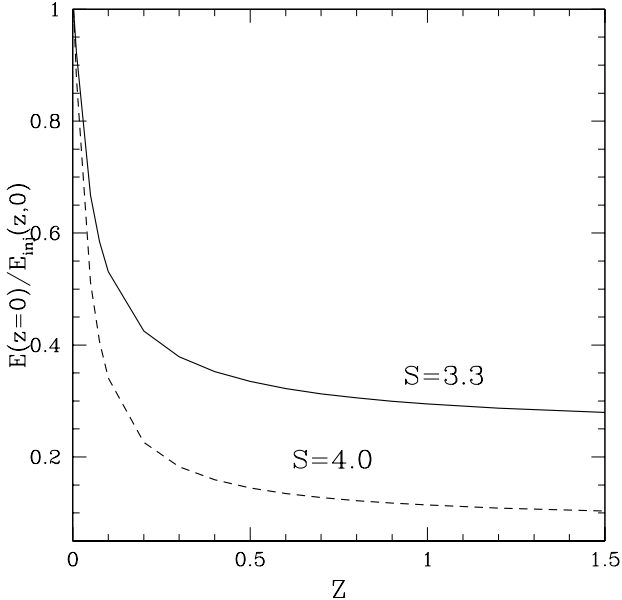}}
\caption[]{The ratio between the energy density of cosmic ray 
protons with momentum $p > 0.1 \, m_p c$ at the present time
($z=0$) 
and the energy density in cosmic rays injected from the redshift
$z_i=z$ (x-axis) to the present.
Calculations are carried out for $s=3.3$ (solid line) and $s=4.0$ 
(dashed line) and for $n_{th}=10^{-3}$cm$^{-3}$.
}
\end{figure}

\begin{figure}
\resizebox{\hsize}{!}{\includegraphics{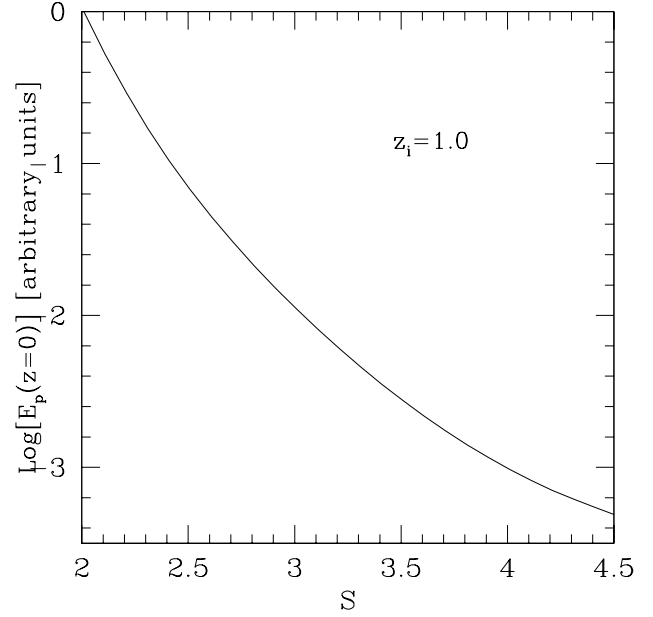}}
\caption[]{Energy density in the form of cosmic ray protons in 
the ICM (arbitrary units) if protons are injected at a constant 
rate (number of particles per unit time) at $p > 0.1 \, m_p c$
with a power law injection spectrum with slope $s$ (on the x-axis).
The time at which injection starts is taken as $z_i=1$. The density
of the thermal gas is assumed to be $n_{th}=10^{-3}$cm$^{-3}$.
}
\end{figure}

Below the threshold for pion production in the $pp$ collisions
(or simply by neglecting the effect due to $pp$ collisions),
the time evolution of the proton spectrum is described by the following 
equation:
\begin{equation}
{{\partial N_{\rm p}(p,t)}\over{\partial t}}=
{{\partial }\over{\partial p}}
\Big(
{ {{ d p }\over{dt}} }_i
N_{\rm p}(p,t) \Big) +Q_{\rm p}(p),
\label{cassano}
\end{equation}
where ${ d p/dt }_i$ is given by Eq. (\ref{coulomb_p}).
Adopting the non-relativistic asymptotic behaviour of Eq.~\ref{propc}, the 
asymptotic solution (for time independent injection) is as follows:
\begin{equation}
N_{\rm p}(p,T)=
{{ K_{\rm p} }\over{n_{\rm th}}}
{{ p^{-s} }\over{ {\cal C} }}
\left\lbrace \begin{array}{llll}
\beta > x_{\rm m} \\
{{ p^3/m_{\rm p}^2 c^2 }\over{
 s-1 }}
\Big\{
1 -  \Big(1+ {{ 3  {\cal C} n_{\rm th} T}\over{p^3/m_{\rm p}^2 c^2 
}} \Big)
^{- {{s-1}\over{3}} }
\Big\} \\
   \\
\beta < x_{\rm m} \\
{{ x_{\rm m}^3 m_{\rm p} c }\over{
s-1}}
\Big\{
1-
\exp( 
{{ n_{\rm th} (1-s) T {\cal C} }\over
{ x_{\rm m}^3 m_{\rm p} c }}
) \Big\} 
\end{array}
\right.
\label{inj_p}
\end{equation}
\noindent
where ${\cal C}$ is the constant in Eq. (\ref{coulomb_p}) and $T$ is the time 
elapsed from the beginning of the injection.
Eq. (\ref{inj_p}) presents the following 
behaviours :
\begin{equation}
N_{\rm p}(p,T) \propto
\cases{
p^{3-s} & if $p<<p_*(T)$ \& $p > m_{\rm p} c x_{\rm m}$ \cr
p^{-s},& if $p<<p_*(T)$ \& $p < m_{\rm p} c x_{\rm m}$ \cr
p^{-s},& if $p>>p_*(T)$; \cr}
\label{asintotip}
\end{equation}

\noindent
where $p_*(T)$ is given by \footnote{Note that in the ultra--relativistic
limit Coulomb losses are negligible and one has $N_{\rm p}(p,t)
\propto p^{-s}$} :

\begin{equation}
p_*(T) \sim 0.7 \times \Big(
{{ T }\over{ 10 {\rm Gyr} }}
{{ n_{\rm th} }\over{ 10^{-3} }}
\Big)^{1/3} m_p c
\label{p*}
\end{equation}

In Fig. 2 we plot the spectrum of the protons measured at the present time
as obtained solving Eq. (\ref{cassano}) numerically under the assumption of a
continuous time-independent injection of protons (starting from different 
$z_i$, see caption) in the cluster volume. The proton spectra, largely 
modified in the trans--relativistic regime by Coulomb losses, has the 
asymptotic behaviour described by Eq. (\ref{asintotip}).

Since the energy flux of Alfv\'en waves, as we show below, is efficiently 
damped on protons, it is important to estimate the amount 
of energy stored in the form of protons in the ICM.

In Fig. 3 we plot the ratio between the energy injected in cosmic ray 
protons (assuming time independent injection starting at redshift $z_i$)  
and the energy stored in the form of supra-thermal protons (here $p > 0.1 
\times m_p c$) at the present time for different values of the
injected spectral index $s$ (see caption). As expected, the ratio between 
the energy stored in protons at the present time and the 
injected energy is smaller for steeper injection spectra, 
due to the effect of Coulomb losses.

Finally, by assuming that a fixed fraction of the thermal protons 
in the cluster volume is accelerated to higher energies at a constant 
rate (starting from $z_i = 1.0$), in Fig. 4 we plot the energy stored 
in the cluster at the present time as a function of the slope of the 
injected spectra $s$.

\section{Alfv\'enic acceleration of relativistic particles}

Alfv\'en waves efficiently accelerate relativistic particles via resonant 
interaction. The condition for resonance between a wave of frequency $\omega$ 
and wavenumber projected along the magnetic field $k_{\Vert}$, and a 
particle of type $\alpha$ with energy $E_{\alpha}$ and projected velocity
$v_{\Vert}=v \mu$ is (Melrose 1968; Eilek 1979):
\begin{equation}
\omega
- \nu {{\Omega_{\alpha}}\over
{\gamma}} - k_{\Vert} v_{\Vert}
=0
\label{resonance}
\end{equation}
\noindent
where, in the quasi parallel case ($k_{\perp} << m_{\alpha}\Omega_{\alpha}/p$),
$\nu=-1$ and $\nu=1$ for electrons and protons respectively.

The dispersion relation for Alfv\'en waves in an isotropic plasma with both 
thermal and relativistic particles can be written as (Barnes \& Scargle, 1973):

\begin{equation}
\omega^2= 
{{
k_{\Vert}^2 v_{\rm A}^2 (1+ \eta_1) }\over{
1 + {4\over 3} \eta_2 + v_{\rm A}^2 
c^{-2} ( 1 + \eta_1) 
}}
\label{dispersion}
\end{equation}

\noindent
where

\begin{equation}
\eta_1=
{{ n_{\rm p,rel} }\over
{n_{\rm th}}} +
{{ m_{\rm e} }\over{
m_{\rm p} }}
\Big( 1 + {{ n_{\rm p,rel} }\over
{n_{\rm th}}}
\Big)
\label{eta1}
\end{equation}

and

\begin{eqnarray}
\eta_2=
{{
n_{\rm e,rel} m_{\rm e} < \gamma_{\rm e} >
+ n_{\rm p,rel} m_{\rm p}
< \gamma_{\rm p} >
}\over{
n_{\rm th} m_{\rm p} }}
\label{eta2}
\end{eqnarray}

Since the number density (and possibly the energy density as well) of the 
thermal component in the ICM is considerably
larger than the corresponding non-thermal 
component, one can show that $\eta_1<< 1$ and $\eta_2<< 1$ so that the 
dispersion relation 
[Eq. (\ref{dispersion})] becomes $\omega \simeq |k_{\Vert}| 
v_{\rm A}$.
Combining the dispersion relation of the waves with the resonant condition, 
Eq. (\ref{resonance}), one can derive the resonant wavenumber, $k_{res}$, for 
a given momentum ($p=m v \gamma$) and pitch angle cosine ($\mu$) of the 
particles:

\begin{equation}
k_{res} \sim | k_{\Vert} | 
= {{ \Omega m}\over{p}}
{ 1 \over
{\Big( \mu \pm {{v_{\rm A}}\over{v}} \Big) }},
\label{kres}
\end{equation}
where the upper and lower signs refer to protons and electrons respectively.
In an isotropic distribution of waves and particles, the particle diffusion 
coefficient in momentum space is given by (Eilek \& Henriksen, 1984):
\begin{equation}
D_{\rm pp}(p,t) =
{{ 2 \pi^2 e^2 v_{\rm A}^2}\over{c^3}}
\int_{k_{min}}^{k_{max}}
{{ W_k(t)}\over{k}}
\Big[
1 -
\Big(
{{v_{\rm A}}\over{c}} \mp
{{ \Omega m }\over{p k}} \Big)^2
\Big] dk,
\label{dpp}
\end{equation}
where the minimum wavenumber (maximum scale length) of the waves interacting 
with particles is given by :
\begin{equation}
k_{min} = {{ \Omega m}\over{p}}
{ 1 \over
{\Big( 1 \pm {{v_{\rm A}}\over{v}} \Big) }}
\label{kmin}
\end{equation}
and $k_{max}$ is given by the largest wavenumber of the 
Alfv\'en waves, limited 
by the fact that the frequency of the waves cannot exceed the proton cyclotron
frequency, namely $\omega < \Omega_p$. It follows that $k_{max} \sim \Omega_p/
v_A$ or $k_{max} \sim \Omega_p/v_M$, $v_M$ being the magnetosonic velocity
(here we assume $k_{max} \sim \Omega_p/v_M$).

In the simple case of a power law spectrum of the MHD waves, 
$W(k) \propto k^{-w}$ (for $k_o < k < k_{\rm max}$), 
Eq. (\ref{dpp}) would give:
\begin{eqnarray}
D_{\rm pp}(p) =
A_w
(\delta B)^2
v_A^2
\left({{ p }\over{B}} \right)^w
\Big(1 \pm {{ v_A }\over{c}}
\Big)^w \times \nonumber\\
\left\{
{1\over{w}}
\pm
{{v_A/c}\over{w+1}}
-{{
(v_A/c)^2}\over
{w(w+1)}}
\right\},
\label{dpp-pl}
\end{eqnarray}
\noindent
where
\begin{equation}
\delta B^2
= 8 \pi 
\int W_k dk,
\end{equation}
\noindent
and
\begin{equation}
A_w
= {{ \pi }\over {2}}
{{ e^{2-w} }\over
{c^{3-w} }}
{{ w-1 }\over {w+2}}
k_o^{w-1}
\label{aw}
\end{equation}

The first order expansion (for $v_A/c<< 1$) of Eq. \ref{dpp-pl} is the 
diffusion coefficient generally used in most recent theoretical papers 
on electron acceleration in galaxy clusters (Ohno et al. 2002; Fujita et al. 
2003).

\subsubsection{Electrons}

From Eq. (\ref{kres}), one can see that the momentum of the electrons which 
can resonate with waves with a given wavenumber $k$ depends on the pitch angle
cosine $\mu$. This resonant momentum can be written as
\begin{equation}
p= {{ \Omega_e m_e}\over{k}}
{1\over{ \mu - {{v_A}\over{v}}}
}.
\label{res_e}
\end{equation}

The minimum momentum of the electrons for which resonance with waves of a 
given wavenumber $k$ can occur is:

\begin{equation}
p_{min} = 
{{ m_e}\over{k}}
\Big( \Omega_e + v_A k \Big),
\label{res_eth}
\end{equation}
\noindent
which, in the relativistic limit becomes:
\begin{equation}
p_{min}^{rel}= {{ \Omega_e m_e}\over{k}}
{1\over{ 1 - {{v_A}\over{c}}}}.
\label{res_erel}
\end{equation}

Since the wavenumber of Alfv\'en waves in a plasma is limited by 
$\omega < \Omega_p$, from Eq. (\ref{res_eth}), one has that the minimum
momentum of the electrons which can resonate with Alfv\'en waves is:

\begin{equation}
p_{min} = p_{th} {{v_A}\over{v_{th}}}
\Big( {{ m_p }\over {m_e}} +1 \Big),
\label{min_eth}
\end{equation}
\noindent
which, in general, gives $p_{min} >> p_{th}$, $p_{th}=m_e v_{th}$ being the 
momentum of the thermal electrons. It follows the well known result that 
thermal electrons cannot resonate with Alfv\'en waves (Hamilton \& Petrosian, 
1992 and references therein).

This important limitation of Alfv\'en waves as particle accelerators forces
us to consider the situation in which a relic population of relativistic
electrons exists in the ICM (Sect. 2.2), 
and no electron acceleration from the thermal 
background is taken into account.

In the case of an isotropic, homogeneus phase--space density electron plasma, 
the evolution of the electron spectrum can be described in the context of the
Fokker--Planck equation (Tsytovich 1966; Borovsky \& Eilek 1986):
\begin{eqnarray}
{{\partial f(p,t)}\over{\partial t}}=
{1\over{p^2}} {{\partial }\over{\partial p}}
\Big[ D_{\rm pp} p^2 {{\partial f(p,t)}\over{\partial p}}
+{{dp}\over{dt}}_{\rm rad} p^2 f(p) + 
\nonumber\\
 + {{dp}\over{dt}}_{\rm i} p^2 f(p) \Big]+
 q^+_e(p,t),
\label{fokker-e}
\end{eqnarray}
\noindent
where $f(p,t)$ is the phase space density of electrons, $D_{\rm pp}$ is the 
diffusion coefficient due to the interaction with the waves [Eq. (\ref{dpp})], 
${dp/dt}_{\rm i}$ and ${dp/dt}_{\rm rad}$ give the ionization and radiative  
losses [Eq. (\ref{ion}-\ref{syn+ic})], and $q^+_e(p,t)$ is an isotropic 
phase--space electron source term. For simplicity we can introduce the 
two functions $N(p,t)$ and $Q_e(p,t)$, related to $f$ and $q^+_e$ through
the following relations:

\begin{equation}
N(p,t)= 4 \pi p^2 f(p,t),
\end{equation}
\noindent
and
\begin{equation}
Q_e(p,t)= 4 \pi p^2 [q^+_e(p,t)].
\end{equation}
\noindent

The diffusion equation Eq. (\ref{fokker-e}) in momentum space can therefore
be transformed into an equation that describes the evolution of the electron 
number density:
\begin{eqnarray}
{{\partial N(p,t)}\over{\partial t}}=
{{\partial }\over{\partial p}}
\left[
N(p,t)\left(
{{dp}\over{dt}}_{\rm rad} + {{dp}\over{dt}}_{\rm i}
-{2\over{p}} D_{\rm pp}
\right)\right] +
\nonumber\\
{{\partial }\over{\partial p}}
\left[
D_{\rm pp}
{{\partial N(p,t)}\over{\partial p}}
\right] +
Q_e(p,t).
\label{elettroni}
\end{eqnarray}

\subsubsection{Protons}

From Eq. (\ref{kres}) one has that the momentum of the protons which may 
resonate with waves with a given wavenumber $k$ is:

\begin{equation}
p= {{ \Omega_p m_p}\over{k}}
{1\over{ \mu + {{v_A}\over{v}}}
}.
\label{res_p}
\end{equation}

The minimum momentum of the protons which may resonate with waves having 
wavenumber $k$ is therefore:

\begin{equation}
p_{min} = 
{{ m_p}\over{k}}
\Big( \Omega_p - v_A k \Big).
\label{res_pth}
\end{equation}
In the relativistic limit this reduces to:

\begin{equation}
p_{min}^{rel}= {{ \Omega_p m_p}\over{k}}
{1\over{ 1 + {{v_A}\over{c}}}}.
\label{res_prel}
\end{equation}

\noindent
From Eq. (\ref{res_pth}), one has that the minimum momentum of the protons 
which can resonate with Alfv\'en waves in units of the momentum of the thermal 
protons is :
\begin{equation}
p_{min} = p_{th} {{v_A}\over{v_{th}}}
\Big( {{ \Omega_p }\over{\omega}} -1 \Big).
\label{min_pth}
\end{equation}
\noindent
Since $\omega < \Omega_p$, this basically means that thermal protons 
can efficiently resonate with Alfv\'en waves (Hamilton \& Petrosian, 1992).

As in the case of relativistic electrons, the evolution of the proton spectrum
is obtained from the equation:
\begin{eqnarray}
{{\partial N(p,t)}\over{\partial t}}=
{{\partial }\over{\partial p}}
\left[
N(p,t)\left( {{dp}\over{dt}}_{\rm i}
-{2\over{p}} D_{\rm pp}
\right)\right] +
\nonumber\\
{{\partial }\over{\partial p}}
\left[
D_{\rm pp}
{{\partial N(p,t)}\over{\partial p}}
\right] +
Q_p(p,t),
\label{protoni}
\end{eqnarray}
where ${dp/dt}_{\rm i}$ is given by Eq.~(\ref{coulomb_p}) and the coefficient
$D_{pp}$ is provided in Eq. (\ref{dpp}).

\subsection{From fluid turbulence to Alfv\'en waves}

\subsubsection{Injection}

We assume that fluid turbulence is present in the cluster volume with a power 
spectrum
\begin{equation}
W_{\rm f}(x_{\rm f})=
W_{\rm f}^{\rm o} x_{\rm f}^{-{\rm m}}
\label{wfluid}
\end{equation}
\noindent
in the range $x_{\rm f}^{\rm min} < x_{\rm f} < x_{\rm f}^{\rm max}$,
where $x_{\rm f}^{\rm min}$ is the wavenumber corresponding to the maximum 
scale of injection of the turbulence and the maximum wavenumber is that at 
which the effect of fluid viscosity starts to be important and it is 
of the order of $x_{\rm f}^{\rm max} \sim x_{\rm f}^{\rm min} 
( {\cal R} )^{-3/4}$ (e.g., Landau \& Lifshitz, 1959), ${\cal R}$ being the 
Reynolds' number.

For Kolmogorov turbulence we have ${\rm m}=5/3$, while for 
Kraichnan turbulence (Kraichnan 1965) one has 
${\rm m}=3/2$ (see Sect.~3.1.2 for a discussion on the Kolmogorov
and Kraichnan phenomenology).

Here we investigate the connection between the fluid turbulence that we 
start with and the MHD waves that we use as particle accelerator.
Fluid turbulence can radiate MHD modes (Kato 1968) via the Lighthill process. 
A fluid eddy may be thought of as radiating MHD waves in the mode $j$ at a 
wavenumber $k = (v_{\rm f}(x)/v_{j})x_{\rm f}$, where $v_{j}$ is the velocity 
of the $j$--mode wave. The MHD modes are expected to be driven only for 
$x > x_{\rm T}$, $x_{\rm T}$ being the wavenumber at which the transition
from large--scale ordered turbulence to small--scale disordered turbulence
occurs. Following previous works in the literature (Eilek \& Henriksen 1984; 
Fujita et al., 2003), we adopt the Taylor wavenumber as an estimate of this
transition scale, namely:

\begin{eqnarray}
l_{\rm T} = {{2 \pi}\over{x_{\rm T}}} 
\sim  
\Big[ < v_{{\rm f,i} }^2 > / < \Big( {{
\partial v_{{\rm f},i} }
\over{ \partial x_i }} \Big)^2 >
\Big]^{1/2}
\nonumber\\
\sim 
l_o (15 / {\cal R} )^{1/2},
\label{taylor}
\end{eqnarray}
\noindent
where the Reynolds number is given by ${\cal R} =l_o v_{\rm f}/\nu_{\rm K}$,
and $\nu_{\rm K}$ is the kinetic viscosity. The fraction of the fluid 
turbulence radiated is small for all but the larger eddies, near the Taylor
scale, and thus the {\it Lighthill} radiation can be expected to not disrupt 
the fluid spectrum. More specifically, the energy rate radiated via the 
{\it Lighthill} mechanism into waves of mode $j$ and wavenumber $k$ is given 
by (e.g., Eilek \& Henriksen 1984):

\begin{equation}
I_{j}(k)=I_{j, {\rm o}}
\Big( {{k}\over{x_{\rm T}}} \Big)^{-y_{j}},
\label{imhd}
\end{equation}
where
\begin{equation}
y_{j}=(2 n_{j}+3)
{{ m -1 }\over{3 -m}} \,
\label{ymhd}
\end{equation}


\begin{equation}
I_{j, {\rm o}}
\sim
\Big|
{{ 2 n_{j} (1-m) +6 -4 m}\over
{3 -m}} \Big|
\rho v_{j}^3 
\Big(
{{
{\cal E}_t
}\over
{\rho v_{j}^2 R}}
\Big)^{ {{ 3+2n_{j} }\over{
3-m}} },
\label{ialpha}
\end{equation}

${\cal E}_t \sim \rho v_f^2$ is the energy density of the
fluid turbulence, and

\begin{equation}
R=
{{ x_o W_f(x_o)}\over
{x_T W_f(x_T)}}
\label{r}
\end{equation}

Here, $n_{j}$=0 for Alfv\'en and slow magnetosonic (MS) waves, and $n_{j}=1,2$ 
for fast MS waves. In the following we concentrate on Alfv\'en waves, in which 
case $y_{j}=3/2$ ($y_{j}=1$) for a Kolmogorov (Kraichnan) spectrum of the 
fluid turbulence. A more general treatment, including the effect of
fast magnetosonic waves, will be given in a forthcoming paper.

\subsubsection{Basic equations and time evolution}

In our calculations we assume for simplicity that Alfv\'en waves propagate
isotropically in the cluster volume and we assume $k \simeq |k_{\Vert}|$.
The spectrum of Alfv\'en waves driven by the fluid turbulence evolves as a
result of wave--wave and wave--particle coupling. In particular, the 
wave--particle involves the thermal and relativistic particles, in the
way explained in the previous Section.
The combination of these processes produces a modified, time--dependent 
spectrum of Alfv\'en waves, $W_{\rm k}(t)$, which can be calculated by solving 
the continuity equation (e.g., Eilek 1979):

\begin{equation}
{{\partial W_{\rm k}(t)}\over
{\partial t}} =
{{\partial}\over{\partial k}}
\left( D_{\rm kk} {{\partial W_{\rm k}(t)}\over{\partial k}}
\right)
-\sum_{i=1}^n \Gamma^{\rm i}_{\rm k} W_{\rm k}(t) +
I_{\rm k}(t).
\label{turbulence}
\end{equation}

\noindent
The first term on the right hand describes the wave--wave interaction, with
diffusion coefficient $D_{\rm kk} = k^2/\tau_s$. $\tau_s$ is the spectral 
energy transfer time, that for a given wavelength is given by $\tau_s \sim
\tau_{NL}^2/\tau_3$ (Zhou \& Matthaeus 1990), where $\tau_{NL}=\lambda/
\delta v$ is the non--linear eddy--turnover time ($\delta v$ is the rms 
velocity fluctuation at $\lambda$) and $\tau_3$ is the time over which this 
fluctuation interacts with other fluctuations of similar size.

In the framework of the Kolmogorov phenomenology, the Alfv\'en crossing time 
$\tau_{A}=\lambda/v_A$ largely exceeds $\tau_{NL}$, therefore fluctuations of 
comparable size interact in one turnover time, namely $\tau_3 \sim \tau_{NL}$.
In the Kraichnan phenomenology, $\tau_A << \tau_{NL}$, therefore convection 
limits the duration of an interaction and $\tau_3 \sim \tau_A$.
Since the velocity fluctuation, $\delta v$, is related to the rms wave field, 
$\delta B$, through the relation $\delta v^2 / v_A^2 = \delta B^2 /B^2$, the
diffusion coefficient is given by (Miller \& Roberts 1995):
\begin{equation}
D_{\rm kk} \simeq v_{\rm A} 
\cases
{k^{7/2}
\left( {{ W_{\rm k}(t) }\over{ 2 W_{\rm B}}} \right)^{1/2},&
(Kolmogorov) \cr
k^4 \left( {{ W_{\rm k}(t) }\over{ 2 W_{\rm B}}} \right),&
(Kraichnan) \cr },
\label{dkk}
\end{equation}
\noindent
in the Kolmogorov and Kraichnan phenomenology, respectively.

The second term in Eq. (\ref{turbulence}) describes the damping with the 
relativistic and thermal particles in the ICM. In the case of nearly parallel 
wave propagation ($k_{\perp} << m \Omega/p$, $k\simeq |k_{\Vert}|$) and 
isotropic distribution of particles of type $\alpha$, the cyclotron damping
rate for Alfv\'en waves is as given by Melrose (1968):
\begin{eqnarray}
\Gamma^{\alpha}_{\rm k}(t)=
-{{ 4 \pi^3 e^2 v_{\rm A}^2 }\over
{k c^2}}
\int_{p_{\rm min}}^{p_{\rm max}}
p^2 (1 - \mu_{\alpha}^2 )
{{ \partial f_{\alpha}(p,t) }\over{ \partial p}}
dp =\nonumber\\
{{ \pi^2 e^2 v_{\rm A}^2 }\over
{k c^2}}
\int_{p_{\rm min}}^{p_{\rm max}}
(1 - \mu_{\alpha}^2 )
\left(
2 {{N_{\alpha}(p,t)}\over{p}}
-{{\partial N_{\alpha}(p,t) }\over{ \partial p}}
\right) dp,
\label{damping}
\end{eqnarray}
where, for relativistic particles, one has :
\begin{equation}
\mu_{\alpha}^{\rm rel}=
{{ v_{\rm A} }\over
{c}} \pm {{\Omega_{\alpha} {\rm m}_{\alpha} }\over
{p k}},
\end{equation}
\noindent
while for sub--relativistic particles:
\begin{equation}
\mu_{\alpha}^{\rm th}=
{{v_{\rm A} {\rm m}_{\alpha} }\over
{p}}
\pm
{{\Omega_{\alpha} {\rm m}_{\alpha} }\over
{p k}}.
\end{equation}
\noindent
Here the upper and lower signs are for negative and positive charged 
particles respectively. Lacombe (1977) showed that the damping rate
for isotropic Alfv\'en waves are well within a factor of $\sim 3$ of that 
calculated for nearly parallel wave propagation, therefore we are justified 
to use Eq. (\ref{damping}) in our calculations.

The third term in Eq. (\ref{turbulence}) describes the continuous injection
of Alfv\'en waves as radiated by the fluid turbulence through the {\it Lighthill}
mechanism. From Eq. (\ref{ialpha}) with $n_{j}=0$ one has:
\begin{equation}
I_k
\simeq 2
\Big|
{{ 3 -2 m}\over{3 -m }}
\Big|
\rho
v_{\rm A}^3 
\Big(
{{ v_{\rm f}^2 }\over
{  v_{\rm A}^2
R }}
\Big)^{ 3 \over{3 -m }}
\times 
k^{-3 {{m-1}\over{3-m}}}.
\label{inj_alfven}
\end{equation}

\begin{figure*}
\resizebox{\hsize}{!}{\includegraphics{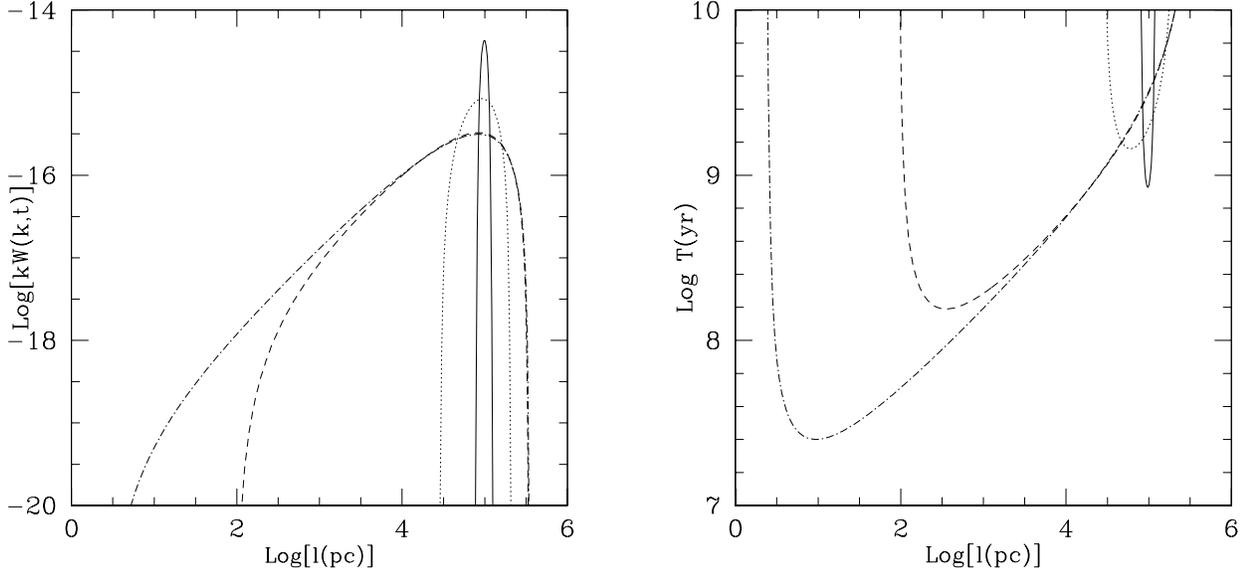}}
\caption[]{
{\bf Left Panel}: Time evolution of the spectrum of Alfv\'en waves
injected in a single burst at a given scale. The spectra are plotted for
$10^{14}$ (solid line), $5\times 10^{15}$ (dotted line),
$3 \times 10^{16}$ (dashed line), and $3.2 \times 10^{16}$ s
(dot-dashed line) after the injection event.
In the calculations, a Kolmogorov diffusion coefficient is adopted.
The temperature of the gas, the magnetic field and the gas density
are $T=10^{8}$K, $B=1 \mu$G, and $n_{th}=10^{-4}$cm$^{-3}$ respectively.
{\bf Right Panel}: Time evolution of the cascade--time scale estimated
for the same parameters and with the corresponding line--styles
as in the Left Panel.}
\end{figure*}

\section{General Results on the evolution of particles and waves}

The interaction between waves and particles is investigated by solving 
the set of coupled differential equations given in Eqs.~(\ref{elettroni}), 
(\ref{protoni}), and (\ref{turbulence}). 
More specifically, the time-evolution of the wave spectrum
depends on the damping processes which are affected by the spectra and energy 
content of the electron and proton components in the ICM. In turn, the 
wave-particle interactions modify the spectra of electrons and protons
while energizing these particles. The various components are clearly 
strongly coupled to each other and they cannot be treated separately as
has been done in previous work on the subject.

In the following we discuss separately in some detail the processes of
turbulent cascading and damping.

\subsection{Turbulent cascade}

The equation that describes the turbulent cascading without accounting for 
damping processes and wave injection is the following:

\begin{equation}
{{ \partial W_k(t)}\over
{\partial t}}
=
{{\partial}\over{\partial k}}
\Big(
D_{kk}
{{ \partial W_k(t)}\over
{\partial k}}
\Big).
\label{cascade}
\end{equation}

The cascade timescale at a given wavelength is $\tau_{\rm kk} \sim k^2/
D_{\rm kk}$, and using Eq.~(\ref{dkk}):
\begin{equation}
\tau_{\rm kk}(l) \simeq
{{ 2 \times 10^8 {\rm yr} }\over
{B_{\mu G} }}
\Big(
{{l_{100}}\over{{\rm kpc}}}
\Big)
\Big(
{{n_{\rm th}}\over{10^{-3}}}
\Big)^{ {1\over2} }
\left\lbrace \begin{array}{lllll}
\sqrt{2} \Big( {{ {\delta B}_{>k}}\over
{B}} \Big)^{-1} \\
{\rm (Kolmogorov)} \\
  \\
2 \Big( {{ {\delta B}_{>k}}\over
{B}} \Big)^{-2} \\
{\rm (Kraichnan)}
\end{array}
\right.
\label{tkk}
\end{equation}
\noindent
where we define ${\delta B}_{>k} \sim \sqrt{ 8 \pi k W_k}$.
It is worth noticing that the cascade timescale in the Kolmogorov regime 
does not depend on the value of the magnetic field strength. We also notice 
that in both Kolmogorov and Kraichnan regimes, the cascade timescale depends 
on the scale of the waves and on the density of the thermal plasma. In 
particular, $\tau_{\rm kk}$ is smaller in low density regions.

In Fig.~5a we plot the evolution of the spectrum of a population of waves 
injected at a given scale as obtained from Eq.~(\ref{cascade}) in the 
Kolmogorov phenomenology (see caption for details); the broadening of the wave
distribution at scales larger than the injection scale clearly shows the 
effect of stochastic wave--wave diffusion. In qualitative agreement with 
Eq. (\ref{tkk}), it is clear from Fig.~5a that the developing rate of 
wave-wave cascade from larger to smaller scales increases with decreasing 
scale.
In Fig.~5b we show the cascade time-scale [from Eq.~(\ref{tkk})] calculated 
for the  spectra (and corresponding times) used in Fig.~5a.
As a general remark, we find that for typical conditions of the ICM, the 
wave--wave time scale below 1 pc, namely on the scale relevant for 
wave--particle interaction, is considerably shorter than $10^7$yr.

\subsection{Damping processes}

The first obvious damping process is provided by the interaction of the 
Alfv\'en waves with the protons in the thermal plasma. Assuming that thermal
particles have a Maxwellian energy distribution with temperature $T$ and 
energy density ${\cal E}_{th}$, the damping rate is obtained from 
Eq.~(\ref{damping}):

\begin{eqnarray}
\Gamma_{\rm p^+_{th}}(k)=
{{
16 \pi^{3/2} e^2 v_A^2 m_{p} {\cal E}_{p,th} }
\over
{3 c^2 p_{\rm th}^3 k}}
\Big[
1+ \Big( {{p_{\rm min}}\over{p_{\rm th}}}
\Big)^2 - \nonumber\\
{{ m_{p}^2 v_A^2 }\over{p_{\rm th}^2}}
\Big( 1+
{{\Omega_{p}^2}\over{k^2 v_A^2}}
- {{2 \Omega_{p}}\over{k v_A}}
\Big) \Big]
{\rm exp}\Big(- ({{p_{\rm min}}\over{p_{\rm th}}})^2
\Big) \nonumber\\
\buildrel {k \sim k_{\Vert}} \over
\longrightarrow
{{
16 \pi^{3/2} e^2 v_A^2 m_{p} {\cal E}_{p,th} }
\over
{3 c^2 p_{\rm th}^3 k}}
{\rm exp}\Big(- ({{p_{\rm min}}\over{p_{\rm th}}})^2
\Big)
\end{eqnarray}
\noindent
where
\begin{equation}
p_{\rm th}
=(2 m_{p} k_B T )^{1/2},
\end{equation}
\noindent
and $p_{\rm min}$ is given by Eqs.(\ref{res_pth}).

It is well known that since the resonant condition [Eq. (\ref{resonance})] 
selects the interaction between particles of momentum $p$ with waves with 
wavenumber $k\propto p^{-1}$, most of the energy of the waves is dissipated 
in the acceleration of relativistic particles (e.g.,
Eilek, 1979), and more 
specifically of relativistic protons. Although the exact damping rate 
with cosmic ray protons used in our calculations is 
obtained by numerically combining Eq.~(\ref{damping}) and the solution of 
Eq.~(\ref{protoni}), it is also 
possible to write an analytical expression for 
the damping rate for the asymptotic solutions obtained for the proton spectrum 
[Eqs.(\ref{inj_p}-\ref{asintotip})]. For supra--thermal protons, 
for $\beta< x_{\rm m}$, the damping rate is given by:

\begin{eqnarray}
\Gamma_{\rm p^+}(k)= 
{{ 2 \pi^2 e^2 m_p }\over{n_{\rm th} c }}
{{ K_p }\over{
{\cal C}}}
{{ v_A^2 }\over{k}}
{{x_{\rm m}^3}\over{s(s-1)}} 
\Big[
{{ \Omega_{\rm p} m_{\rm p} }\over
{k}} \Big(1 - {{ v_A k}\over{\Omega_{\rm p}}}
\Big) \Big]^{-s}
\nonumber\\
\times \Big\{1-
{\rm exp}\Big(
{{ n_{\rm th} (1-s) T {\cal C} }\over
{x_{\rm m}^3 m_p c}}
\Big)  \Big\}, 
\label{damp-pp1}
\end{eqnarray}
\noindent
while, for $\beta > x_{\rm m}$:

\begin{eqnarray}
\Gamma_{\rm p^+}(k)=
-
{{ \pi^2 e^2 }\over{n_{\rm th} (m_p c^2)^2 }}
{{ K_p }\over{
{\cal C}}}
{{ v_A^2 }\over{k}}
\Big\{
\Big(
{{ \Omega_p m_p }\over{k}}
( 1 - {{ v_A k}\over{
\Omega_p}} )
\Big)^{1-s}
\nonumber\\
\times \Big[
-{2 \over{ (s-1)(s-3) }}
\Big( {{\Omega_p m_p }\over{k}}
\Big)^2+ \nonumber\\
2 {{\Omega_p m_p^2 v_A }\over{
k}}[ {1\over{s-3}} - {\cal L}^{-(s-1)/3} ]-
\nonumber\\
m_p^2 v_A^2 \Big[
{1\over{s-3}} + {{ {\cal L}^{-(s-1)/3} }\over
{s-1}} \Big] \Big]+
\nonumber\\
 {2\over{s-1}}
\int_{p_{min}}
p^{2-s} {{\cal L}}^{-(s-1)/3}
dp \Big\}.
\label{damp-pp2}
\end{eqnarray}
\noindent
Here
\begin{equation}
{{\cal L}}=
1 + {{ 3 {\cal C} T n_{\rm th} }\over{p^3/(m_pc)^2}}.
\end{equation}
\noindent 
Finally, in the ultra--relativistic limit we distinguish the two regimes 
$p_{\rm low}> p_{\rm min}(k)$ and $p_{\rm low} \leq p_{\rm min}(k)$, where
$p_{\rm low}$ is the low energy cutoff in the proton distribution
and $p_{\rm min}(k)$ is given by Eq. (\ref{res_prel}).

For $p_{\rm low}> p_{\rm min}(k)$ the damping rate is given by :
\begin{eqnarray}
\Gamma_{\rm p^+}(k)=
{{ \pi^2 e^2 v_A^2 }\over
{c^3}}
{{
{\cal E}_p (s-2)}\over
{p_{low}^2}} {1\over{k}}
\Big[
{{ s +2 }\over{s}}
\Big( 1 - ( {{v_A}\over{c}} )^2
\Big) \nonumber\\
+ {{2 (s+2)}\over{s+1}} {{ v_A}\over{c}}
{{
\Omega_p m_p }\over{
p_{\rm low} k }}
- 
\Big(
{{
\Omega_p m_p }\over{
p_{\rm low} k }}
\Big)^2
\Big],
\label{damp-pp3}
\end{eqnarray}
\noindent
while, for $p_{\rm low} \leq p_{\rm min}(k)$, one has :

\begin{eqnarray}
\Gamma_{\rm p^+}(k)=
{{2 \pi^2  v_A^2 }\over{c}}
\Big(
{{ p_{\rm low}}\over{
\Omega_p m_p }}
\Big)^{s-2}
{{
{\cal E}_p }\over{
B^2}}
\Big[
{{ s-2}\over{s}} +
\nonumber\\
{{ (s+2)(s-2) }\over
{s+1}}
{{ v_A }\over
{c}}
\Big]
k^{s-1}.
\label{damp-pp4}
\end{eqnarray}

The damping rate due to relativistic electrons is obtained combining 
Eq. (\ref{damping}) and the solution of Eq. (\ref{elettroni}).
A first comparison between the strengths of the wave-damping rates  
due to relativistic electrons and protons can be obtained assuming 
a simple power law energy distribution of the relativistic particles.
For $k<< k_{max}$, the resonance occurs for $p_{\rm low} \leq p_{\rm min}(k)$
(i.e., Eq.~(\ref{damp-pp4}) for protons and electrons) for both species and the 
ratio between the proton and electron damping rate is given by :
\begin{equation}
{{
\Gamma_{p^+} }\over{
\Gamma_{e^-}}}
=
\Big(
{{ p^+_{\rm low}}\over
{ p^-_{\rm low}}}
\Big)^{s-2}
{{
{\cal E}_p }\over
{ {\cal E}_e }}
\simeq
{{
N_p^{\rm rel} }\over
{N_e^{\rm rel}}}
\Big(
{{ p_{\rm low}^+ }\over
{p_{\rm low}^-}}
\Big)^{s-1},
\label{dampingevsp}
\end{equation}
\noindent
where the signs $+$ and $-$ refer to the case of protons and electrons 
respectively. From Eq. (\ref{dampingevsp}) it is immediately clear that (for 
a typical $p_{\rm low}^+ / p_{\rm low}^- > 100$, and for $N_p^{\rm rel} \sim 
N_e^{\rm rel}$) the damping rate on protons largely dominates that on
electrons.

\begin{figure}
\resizebox{\hsize}{!}{\includegraphics{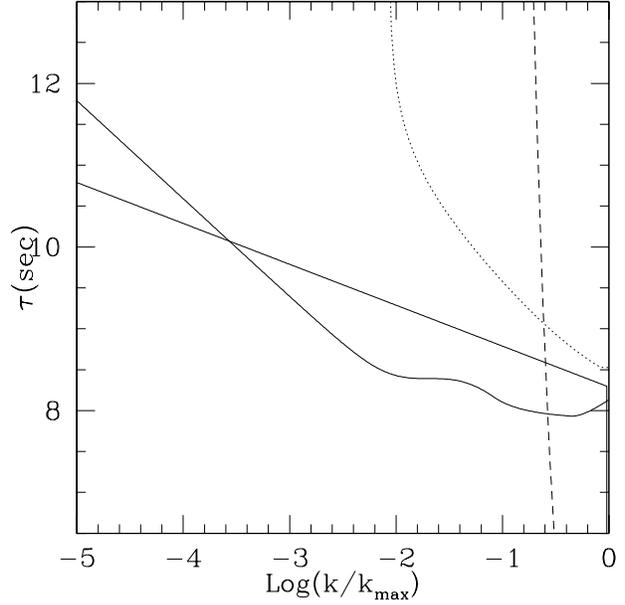}}
\caption[]{
Comparison between damping and cascade time--scales.
The plot shows the time scale for damping on the thermal gas (dashed 
line), on relativistic electrons (dotted line), and on relativistic protons
(solid line). The cascade time scale is plotted as a solid straight line.
Calculations are carried out assuming a Kolmogorov diffusion 
coefficient, and adopting $T=10^8$K, $n_{th}=10^{-3}$cm$^{-3}$,
${\cal E}_{e}=10^{-3}\times {\cal E}_{th}$,
${\cal E}_{p}=10^{-2}\times {\cal E}_{th}$, $s=2.2$ and
$z_i=1.0$, 
$d (\delta B)^2 /d t = 3.3 \times 10^{-15} (\mu {\rm G})^2$/s
and $I_k \propto k^{-3/2}$.}
\end{figure}

\subsection{Damping versus Cascading}

The global damping time can be written as
\begin{equation}
\tau_{\rm d}
= \Big(
\sum_{j=1}^{3} \Gamma_k^{j} 
\Big)^{-1}.
\label{damp-part}
\end{equation}
\noindent
For typical conditions in the ICM, the damping time on the thermal proton
gas is $<10^5$sec (but the process is efficient only for $k/k_{\rm max} > 
0.1$). The damping time on the relativistic component (especially protons) 
is usually $>10^8$sec.

The time scale for the development of the wave-wave cascade depends on the 
wave-wave diffusion coefficient, $D_{kk}$, and thus on the energy density 
of the waves (Sect. 4.1).
Given a spectrum of injection of waves per unit time, $I_k$, one simple 
possibility to estimate the cascade time scale and thus to compare it with 
the time scale of the damping processes is to use the spectum of the waves
under stationary conditions and without damping processes, namely
\begin{equation}
W_k \sim 
{1\over k}
\left\lbrace \begin{array}{lll}
\Big(
{{B^2}\over{4 \pi}}
{{
I_k^2}\over{v_A^2}}
\Big)^{1/3} ,\,
{\rm (Kolmogorov)} \\
\\
\Big(
{{B^2}\over{4 \pi}}
{{ I_k }\over
{ v_A }}
\Big)^{1/2} ,\,  {\rm (Kraichnan)}
\end{array}
\right.
\end{equation}

\noindent
The wave-wave time scale is therefore given by :

\begin{equation}
\tau_s = {{k^2}\over{D_{kk}}}
\sim
{1\over k}
\left\lbrace \begin{array}{lll}
\Big(
{{B^2}\over{4 \pi}}
\Big)^{1/3} /
\Big(
v_A^{2/3}  I_k^{1/3}
\Big) ,\,
{\rm (Kolmogorov)} \\
\\
\Big(
{{B^2}\over{4 \pi}}
\Big)^{1/2}
/
\Big(
v_A^{1/2}  I_k^{1/2}
\Big) ,\,
{\rm (Kraichnan)}
\end{array}
\right.
\label{tau-ww}
\end{equation}

A comparison between the time scales of the damping processes and of the 
wave-wave cascade is given in Fig. 6 for typical values of the parameters 
(assuming a Kolmogorov phenomenology, see caption). 
Fig. 6 shows that the time scale due to the damping with the
thermal pool is considerably shorter than the cascade time scale
for $k/k_{max} >> 0.1$ so that a break or a cutoff in the spectrum of the 
waves is expected at large wavenumbers.
However, the most important result illustrated in Fig. 6 is that, if a 
relatively large number of relativistic protons is present in the ICM, the 
resulting damping time scale can become comparable with or shorter than the 
wave-wave cascade time scale. This means that, at the corresponding 
wavenumbers, the spectrum of the waves is modified by the effect of the 
dampings and therefore that a power law approximation for the spectrum of 
the MHD waves cannot be achieved. We also note that the effect of the 
damping due to relativistic protons is particularly evident at those 
wavenumbers that can exhibit a resonance with the bulk of the relativistic 
electrons in the ICM (those with $\gamma \sim 200-1000$) and thus that this 
effect may have important consequences for the acceleration of the 
relativistic electrons.

\subsubsection{Dependence on the spectrum and energetics of protons}

As already mentioned in the previous Section, the damping of the Alfv\'en 
waves on the relativistic protons modifies the spectrum of the waves 
and therefore indirectly affects the acceleration of electrons. We 
discuss here in some detail the dependence of this damping process
upon the spectrum and energetics of the proton component in the ICM.

The damping of the waves at a given wavenumber basically depends on 
the number of protons with momentum that can resonate with such waves.
At fixed number of relativistic protons with supposedly a power law
spectrum $N(p)\propto p^{-s}$, the damping rate at wavenumbers
corresponding to $p >> p_{\rm low}$ ($p_{\rm low}$ being the minimum 
momentum in the proton spectrum) decreases with increasing $s$.

The efficiency of the damping in galaxy clusters is further
reduced in the case of steep spectra since, in this case, 
Coulomb losses affect the bulk of protons. 

The computed damping time scales, obtained using Eq. (\ref{damp-part}), 
are plotted in Fig. 7 for different proton spectra containing the 
same number of injected protons (see caption): only for large wavenumbers, 
where the condition $p >> p_{\rm low}$ is not applicable and the spectrum
of the resonating protons is not a power law, the damping rate in 
the case of steep spectra is comparable to that obtained for flat 
proton spectra.

\begin{figure}
\resizebox{\hsize}{!}{\includegraphics{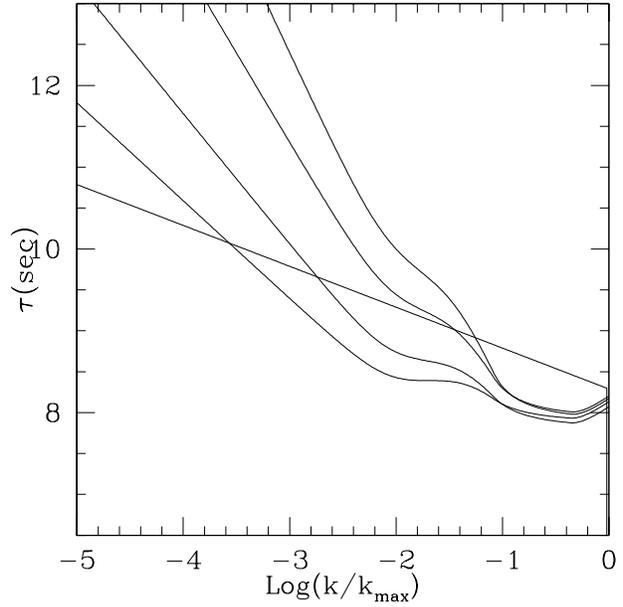}}
\caption[]{
Time scale for damping of Alfv\'en waves on relativistic protons.
From bottom to top the curves show this time scale for the following
choices of the slope of the proton injection spectrum and energy
content (e.g., Fig.4) with respect to the thermal budget:
$s=2.2$ and ${\cal E}_p
= 0.1 \times {\cal E}_{th}$,
$s=2.6$ and ${\cal E}_p
= 0.016 \times {\cal E}_{th}$,
$s=3.2$ and ${\cal E}_p
= 0.002 \times {\cal E}_{th}$,
$s=3.8$ and ${\cal E}_p
= 0.0005 \times {\cal E}_{th}$.
For comparison the cascade time--scale is plotted as a solid straight line.
}
\end{figure}

\subsubsection{Dependence on the physical conditions in the ICM}

The development of the turbulent cascade and the damping of the
waves are affected mainly by the density of the thermal plasma 
$n_{\rm th}$, and by the magnetic field in the ICM.

\begin{itemize}
\item[{\it i)}]
From Eq. (\ref{damping}) and Eqs. (\ref{damp-pp1}--\ref{damp-pp4}), 
at the zeroth order, one has that the damping time scale due to
protons, Eq. (\ref{damp-part}), scales as $\tau_d \propto 
{\cal E}_p^{-1} n_{\rm th}$, while from Eq. (\ref{tau-ww}) one 
has that the time scale for wave-wave cascade is  
$\tau_{kk} \propto n_{\rm th}^{1/3}$. Both these time scales
increase with increasing density of the ICM. However, the damping 
time due to protons is more sensitive to the gas density, in 
a way that, for a given ${\cal E}_p$, the damping processes become
less important for the shape of the spectrum of the waves as
$n_{th}$ increases: the comparison between the two time scales 
for two values of $n_{\rm th}$ is shown in Fig.8. 

If one assumes now that ${\cal E}_p \propto n_{th}$, the 
situation may change: 
in this case the damping time due to protons is constant and the 
importance of the damping processes is minimized in a low density ICM.

\item[{\it ii)}]
The effect of the magnetic field strength on the efficiency of the 
damping processes and on the wave-wave cascade is more complex to 
evaluate, also due to the fact that the maximum wavenumber of the 
MHD waves depends on the magnetic field strength 
(in this paper it is $k_{max}\propto B$).
Assuming a time-independent injection power of MHD waves, $\int I_k 
dk$, from Eq. (\ref{tau-ww}) one has that the time scale for the wave-wave 
cascade is $\tau_{kk} \propto B^{-1/3}$.
Such a simple dependence is not obtained for the damping
processes since from Eq. (\ref{damping}) and Eqs.(\ref{damp-pp1}--
\ref{damp-pp4}) one has that for large values of the wavenumber $k$, 
the damping time scale decreases with increasing $B$, while for 
smaller values of $k$, it is $\tau_d \propto B$.
This is illustrated in Fig.9: the efficiency of the damping 
on ultra--relativistic protons (at small $k$) with respect to the 
turbulent cascade decreases with increasing magnetic field strength.
On the other hand, the opposite trend is seen in the case of mildly
and trans-relativistic protons (i.e., $k/k_{max} \leq 10^{-3}$).
In general, the acceleration of relic ($\gamma < 1000$) relativistic 
electrons is powered by the resonance with waves with relatively 
large values of the wavenumber (e.g., $k/k_{max} \geq 10^{-3}$). 
Thus, given an injection rate of Alfv\'en waves, the results shown in 
Fig.9 indicate that the waves necessary to accelerate electrons 
with $\gamma <10^5$ get appreciably damped by relativistic
protons when the magnetic field strength in the ICM is increased.
The opposite trend is seen for the waves which resonate
with very high energy electrons (e.g., $\gamma > 10^5$).
On the other hand, however, it should be stressed that, in general, 
the efficiency of the wave--particle resonance increases with 
increasing $B$ [Eq. (\ref{dpp}, \ref{dpp-pl})], thus the conclusions 
given in this paragraph do not automatically imply that the electron
acceleration is more efficient in regions with low magnetic field.
\end{itemize}

\begin{figure}
\resizebox{\hsize}{!}{\includegraphics{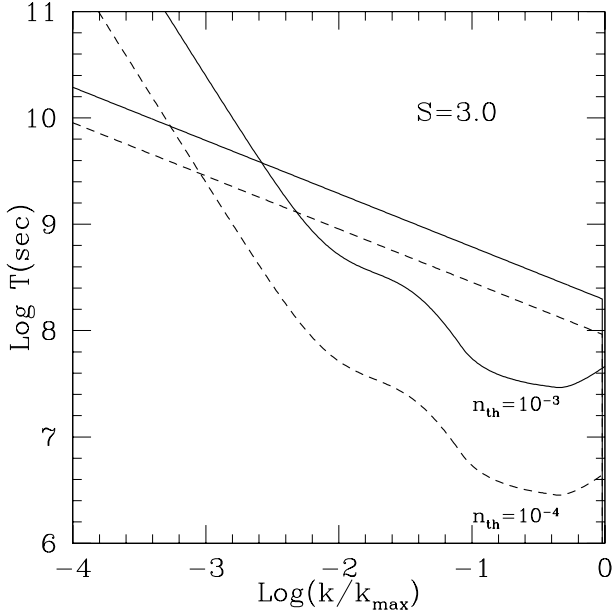}}
\caption[]{
Time scale for damping of Alfv\'en waves on protons as compared 
with the cascade time scale (straight lines) for $n_{th} = 10^{-4}$
cm$^{-3}$ (solid lines) and $10^{-3}$cm$^{-3}$ (dashed lines) and 
for a constant energy density in the form of cosmic ray protons.
The injected spectrum of cosmic ray protons is taken as a power law
with slope $s=3.0$. The injection starts at $z_i=1.0$. The energy 
density is ${\cal E}_p=0.2 \times {\cal E}_{th}$ and $=0.02 \times 
{\cal E}_{th}$ in the low and high density case, respectively.
All the other parameters are fixed as in Fig.~6.}
\end{figure}

\begin{figure}
\resizebox{\hsize}{!}{\includegraphics{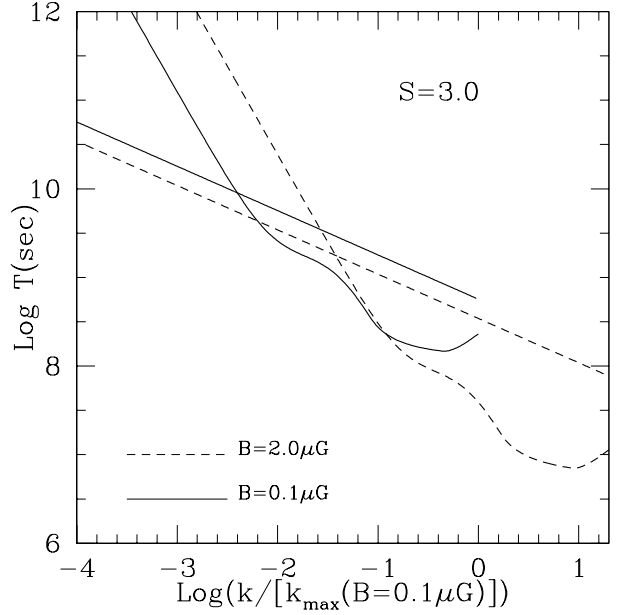}}
\caption[]{
Time scale for damping of Alfv\'en waves on relativistic protons 
compared with the cascade time scale for different values of the
magnetic field strength: $B = 0.1$ (solid lines) and 2.0 $\mu$G 
(dashed lines). It is assumed that $n_{th} = 10^{-3}$cm$^{-3}$, 
$s=3.0$, ${\cal E}_p = 0.02 \times {\cal E}_{th}$. Other
physical parameters being are in Fig.~8. 
}
\end{figure}

\begin{figure}
\resizebox{\hsize}{!}{\includegraphics{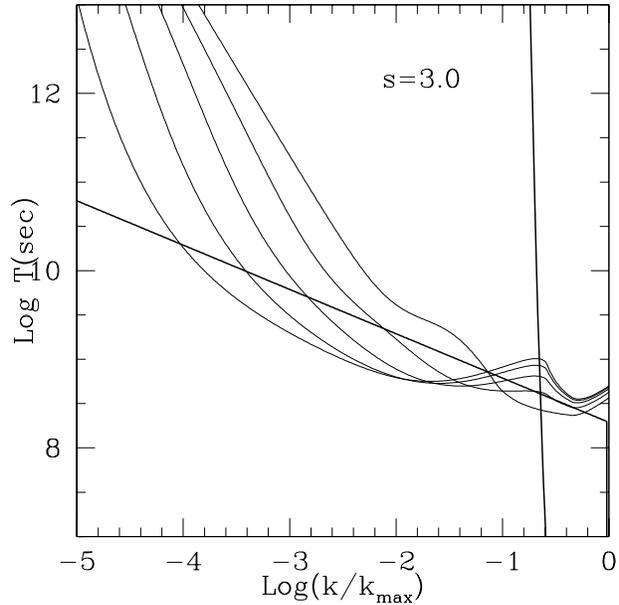}}
\caption[]{
Temporal evolution of the time scale of damping of Alfv\'en waves 
on relativistic protons compared with the cascade time-scale and with
the time scale for damping on the thermal protons (thick curves).
The curves refer to  for 0, $2\times 10^{15}$,
$5\times 10^{15}$, $1\times 10^{16}$sec, and
$2\times 10^{16}$ after the beginning of the acceleration
phase (from the top to the bottom (for $k/k_{\rm max}<0.1$)).
In the calculations we assume :
$d(\delta B)^2/d t = 3.3 \times 10^{-15} (\mu {\rm G})^2$/s,
$T=10^8$K, $n_{th}= 10^{-3}$cm$^{-3}$, $B=0.5 \mu$G,
${\cal E}_e= 0.001 \times {\cal E}_{th}$,
${\cal E}_p= 0.0025 \times {\cal E}_{th}$, $s=3.0$,
$p_{inj} > 0.1 \, m_p c$, and $z_i=1.0$.}
\end{figure}

\begin{figure}
\resizebox{\hsize}{!}{\includegraphics{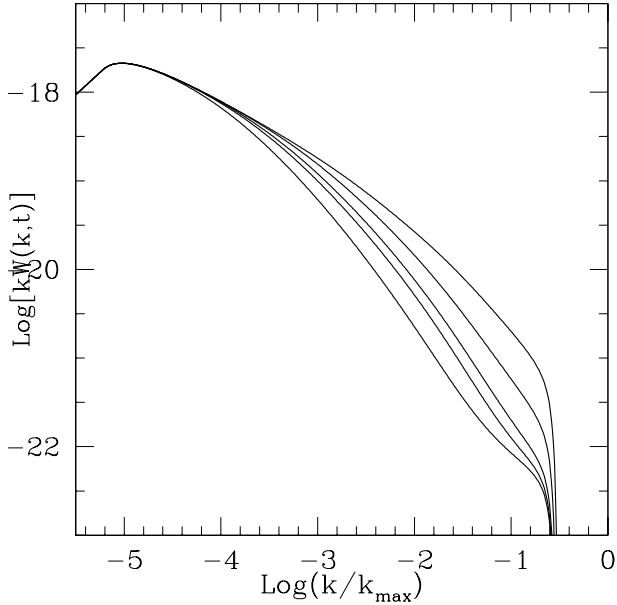}}
\caption[]{
Temporal evolution of the spectrum of the Alfv\'en waves 
at times $2 \times 10^{15}$, $5 \times 10^{15}$, $8 \times 10^{15}$, 
$10^{16}$, and $1.5 \times 10^{16}$sec after the beginning of the 
acceleration (from top to bottom). The calculations are carried out 
for a Kolmogorov spectrum of the fluid turbulence (i.e., $y_j = 3/2$),
a Kolmogorov diffusion coefficient 
$d(\delta B)^2/d t = 3.3 \times 10^{-15} (\mu {\rm G})^2$/s,
$T=10^8$K, $n_{th}= 10^{-3}$cm$^{-3}$, $B=0.5 \mu$G,
${\cal E}_e= 0.001 \times {\cal E}_{th}$,
${\cal E}_p= 0.005 \times {\cal E}_{th}$, $s=3.2$,
$z_i =1.0$ and $p_{inj} > 0.1 \, m_p c$.
The Taylor scale is at $k \sim 10^{-5} k_{\rm max}$.}
\end{figure}

\begin{figure*}
\resizebox{\hsize}{!}{\includegraphics{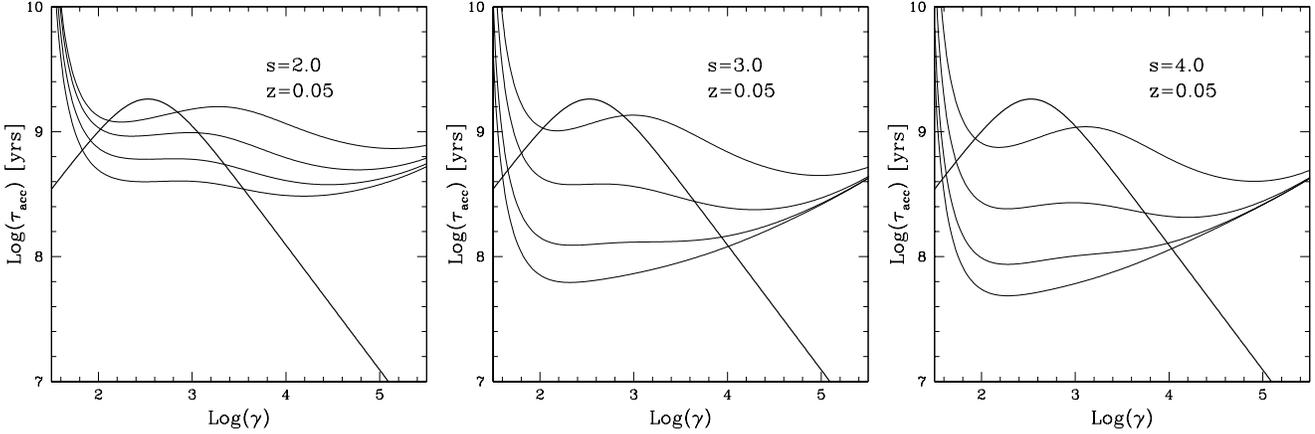}}
\caption[]{
Acceleration time--scale (thin lines) and time--scale for energy losses 
(thick lines) for relativistic electrons as a function of the Lorentz factor.
From bottom to top, the acceleration time--scales are calculated at
$2\times 10^{15}$, $5 \times 10^{15}$, $10^{16}$, and $2 \times 10^{16}$sec 
after the beginning of the acceleration. The following values of the
parameters are adopted:
$d(\delta B)^2/d t = 3.3 \times 10^{-15} (\mu {\rm G})^2$/s,
$T=10^8$K, $n_{th}= 10^{-3}$cm$^{-3}$, $B=0.5 \mu$G,
and ${\cal E}_e= 0.001 \times {\cal E}_{th}$.
{\bf Left Panel}:
${\cal E}_p= 0.2 \times {\cal E}_{th}$, $s=2.0$ and $z_i=1.0$;
{\bf Central Panel}:
${\cal E}_p= 0.025 \times {\cal E}_{th}$, $s=3.0$ and $z_i=1.0$;
{\bf Right Panel}:
${\cal E}_p= 0.002 \times {\cal E}_{th}$, $s=4.0$ and $z_i=1.0$;
}
\end{figure*}

\begin{figure*}
\resizebox{\hsize}{!}{\includegraphics{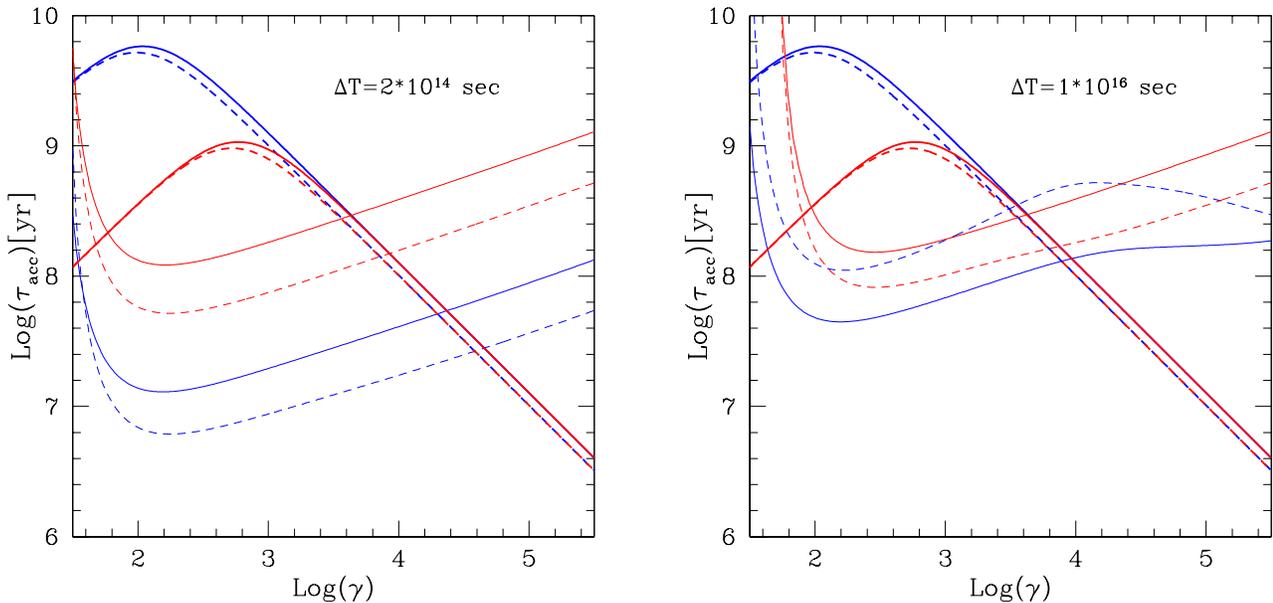}}
\caption[]{
Acceleration time--scales (thin lines) 
and losses time--scales (thick lines) 
for relativistic electrons as a
function of the Lorentz factor for 
$2 \times 10^{14}$sec ({\bf Left Panel}) 
and $10^{16}$sec ({\bf Right Panel}) from the
beginning of the acceleration.
Calculations are shown for $n_{th}=3 \times 10^{-3}$cm$^{-3}$
(red lines) and $n_{th}=10^{-4}$cm$^{-3}$
(blue lines), and for $B=0.1 \mu$G (solid lines)
and $B=1.5 \mu$G (dashed lines).
In the calculations we assumed:
$d(\delta B)^2/dt = 3.3 \times 10^{-15} (\mu{\rm G})^2$/s,
$T=10^8$K, ${\cal E}_e= 0.001 \times {\cal E}_{th}$,
${\cal E}_p= 0.01 \times {\cal E}_{th}$,
and $s=2.2$, not specified parameters being those in
Fig.~12.}
\end{figure*}

\begin{figure}
\resizebox{\hsize}{!}{\includegraphics{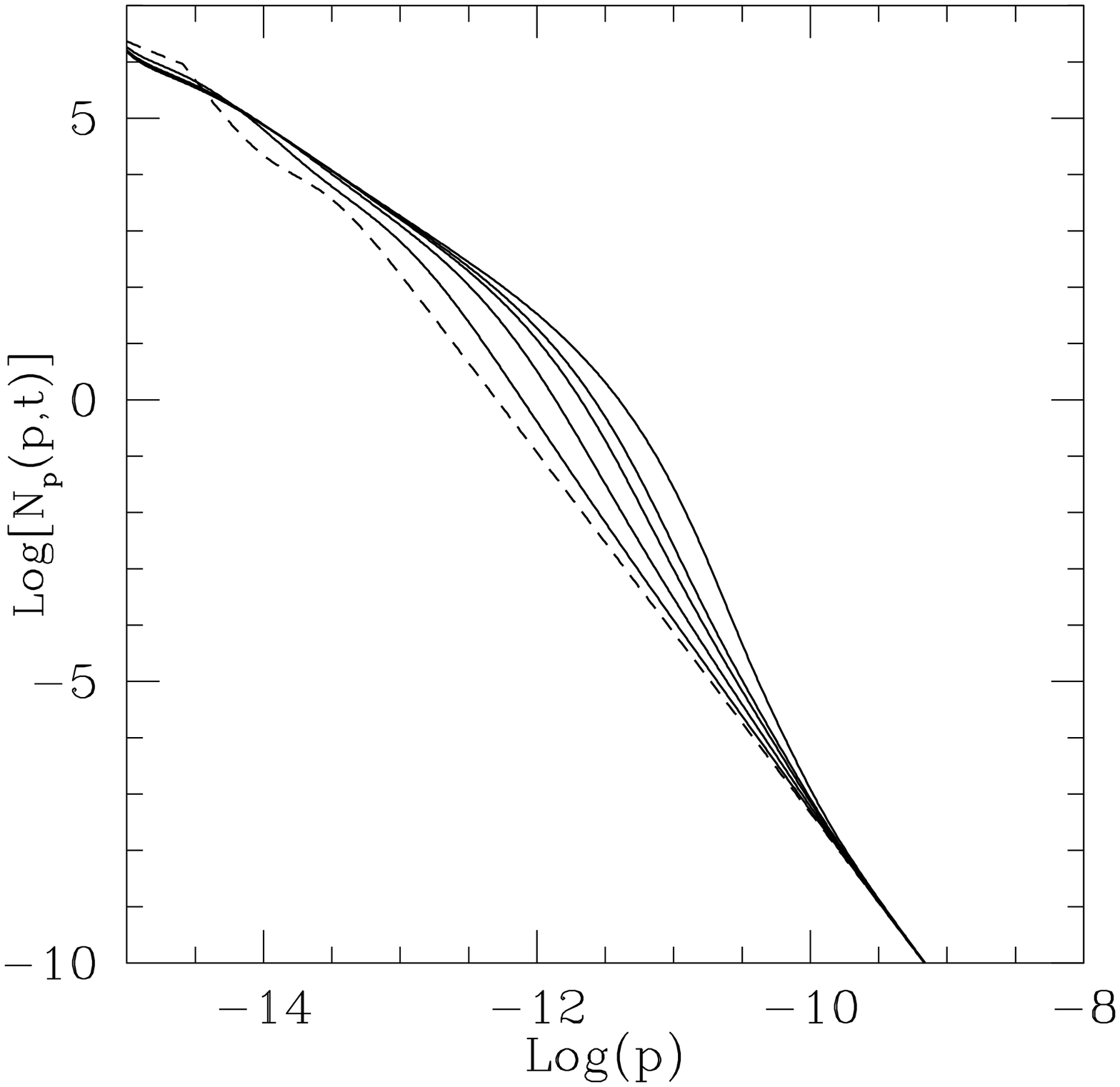}}
\caption[]{
Time evolution of the spectrum of cosmic ray protons as a function of $p$.
From bottom to top, the curves are obtained at times $2\times 10^{15}$, 
$5 \times 10^{15}$, $8 \times 10^{15}$, $10^{16}$, and $1.5 \times 10^{16}$
sec after the beginning of acceleration. The initial proton spectrum is 
plotted as a dashed line. The values of the parameters are as in 
Fig.~11.}
\end{figure}

\section{Quasi Stationary Solutions}

For a given model for the injection of the waves in the ICM, the
spectra of electrons, protons and waves can be calculated using
Eqs.~(\ref{elettroni}) with $Q_e=0$, 
(\ref{protoni}) and (\ref{turbulence}). A few 
comments are in order:

\begin{itemize}
\item[{\it i)}]
In this paper we confine our attention to the case of Alfv\'en waves, 
that resonate predominantly with relativistic particles. Moreover, 
we do not consider situations in which the amount of energy injected 
in the form of turbulence becomes comparable with the thermal energy 
of the ICM. As a consequence, we can safely assume that the thermal 
distributions of electrons and protons in the ICM are not appreciably
affected by the acceleration processes discussed above.

\item[{\it ii)}]
The spectra of electrons, protons and waves, as discussed
above, result from a coupling between all these components:
the spectrum of the waves develops in time due to the turbulent 
cascade until damping becomes efficient and particle acceleration 
occurs. It is worth noticing that the time scales for the processes 
of damping and cascading are quite different from those related to 
particle losses and transport. While the wave spectrum develops 
over $\sim 10^7$ sec, particle acceleration occurs on time scales 
of $\geq 10^{14}$sec, and we are interested in following the 
particle evolution for a typical time of $\geq 10^{15}$sec. 

The clear difference among these time scales suggests that we use 
a {\it quasi stationary approach}, in which it is assumed that at 
each time-step the spectrum of the waves approaches a stationary 
solution (obtained by solving Eq. (\ref{turbulence}) with $\partial W/ 
\partial t =0$) and that this solution changes with time due to the 
evolution of the spectrum of the accelerated electrons and protons.
In other words, at any time-step we solve a set of three differential 
equations: Eq. (\ref{turbulence}) (with $\partial W/ \partial t =0$),
Eq. (\ref{protoni}) and Eq. (\ref{elettroni}).
Intermittent injection of turbulence in the ICM may occur on 
time--scales $\geq 10^7-10^8$ yrs
which are much longer than the time--scales of damping and cascading,
and thus the quasi--stationary 
approach discussed above remains applicable.

\end{itemize}

\subsection{The spectrum of Alfv\'en Waves}

The shape of the spectrum of waves at any time is determined
by the damping of these waves, mainly on protons. The proton 
spectrum in turn changes because of acceleration, and backreacts
upon the spectrum of waves: this implies that even for a 
time-independent rate of injection of waves, the strength of 
the damping rates and the spectrum of the MHD waves are
expected to change with time.

In Fig. 10 we plot an example of the evolution of the time scale 
for the damping due to relativistic protons: it is clear how 
the damping rate increases with time, as a consequence of the
fact that most of the energy injected in MHD waves is channelled 
into relativistic protons. 

A relevant example of the time evolution of the spectrum of waves
is illustrated in Fig. 11: as expected, the energy associated with 
MHD waves which contribute to the acceleration of the bulk of the
relativistic electrons ($k/k_{max}\sim 10^{-1}-10^{-3}$) decreases 
with time. In addition to the general finding that the spectrum of 
the Alfv\'en waves evolves with time, here we also point out that :

\begin{itemize}

\item[{\it a)}]
the spectrum is not a simple power law: it has a different slope
at different $k$ and the curvature of the spectrum also changes 
with time;

\item[{\it b)}]
the spectrum has a low-$k$ cutoff due to the maximum injection
scale, close to the Taylor scale;

\item[{\it c)}]
the spectrum has a high-$k$ cutoff generated by the damping with 
the thermal particles.

\end{itemize}

\subsection{Electron acceleration}

The process of electron acceleration via Alfv\'en waves has been extensively  
investigated in the literature. Eilek \& Henriksen (1984) showed 
that a self-similar solution  for the spectra of electrons and waves 
can be found if these spectra are both required to be power laws
in energy and wavenumber respectively. In this case, the slopes of
the electron ($\delta$) and of the waves ($\omega$) spectrum are 
related by $\delta = 6 - \omega$.

As pointed out above, the assumption of power law spectra is usually
not fulfilled. In fact, more often a stationary solution is found in
the form of a {\it pile up} spectrum $N(p) \propto p^2 \exp\{-p/p_c\}$
(e.g., Borowsky \& Eilek, 1986).

In the general case considered here, the electron spectrum may
be even more complex due to the fact that the spectrum of the waves 
is not a power law and the whole evolution is time-dependent. Note
that the initial stage of reacceleration of relic relativistic
electrons (i.e. $\gamma \sim 100-1000$ electrons) is mainly 
affected by the competition between Coulomb losses and acceleration 
due to the Alfv\'en waves, while later stages, of further acceleration
to the highest allowed energies, are limited by radiative losses.

In Fig. 12 we plot the time scale for electron acceleration, $\sim 
p^2/2 D_{pp}$, and compare it with  the time scale of the energy 
losses of electrons for different proton spectra injected in the 
ICM (see caption). For steep proton spectra, the electron acceleration
is more efficient because less energy gets channelled into the 
proton component. In general, hard proton spectra make the 
acceleration of electrons to Lorentz factors $\gamma > 10^3$
relatively difficult.

If this is true in the initial stage of evolution of the system, 
it becomes increasingly less so at later times: after about 
$\sim 0.5-0.7$ Gyr, relativistic protons have accumulated
enough of the waves energy that the damping of the waves becomes
even more efficient and further acceleration of electrons is
prevented.

In Fig.~13 we plot the acceleration and cooling time scales of 
electrons as computed for different values of the density of 
the ICM and of the magnetic field strength.

It is clear that at the beginning of the reacceleration phase
(Fig.~13a) the electron acceleration is enhanced by 
increasing the magnetic field strength and by decreasing
the density of the ICM. On the other hand, at later times
in the reacceleration phase, the situation can be much more
complicated. In particular, when the reacceleration efficiency 
is very high (for instance for large values of the magnetic 
field strength and for low values of $n_{th}$), the energy 
stored in relativistic protons after few reacceleration times 
may become sufficiently high to increase the damping rate of 
the Alfv\'en waves and thus decrease the efficiency of electron 
acceleration (Fig.~13b): in this case, higher values of the 
magnetic field strength produce a lower acceleration efficiency 
compared with the case of low magnetic field.

The continuous backreaction between waves and protons creates
a sort of {\it wave-proton boiler} that in a way is 
self-regulated.

If the injection of fluid turbulence is intermittent on time 
scales of the order of the cooling time of electrons with Lorentz factors
$\gamma \sim 10^3-10^4$, then the effect of the {\it wave-proton boiler} 
on the electron acceleration may be reduced. The reason for this is that
for a given reacceleration rate, the accumulation of energy in the form 
of relativistic protons requires longer times and the electron acceleration 
remains efficient for $\sim 1$ Gyr.

\subsection{Proton acceleration}

Alfv\'enic acceleration of thermal and relativistic protons 
is extensively studied in the literature, in particular as
applied to the case of solar flares (e.g., Miller, Guessoum, 
Ramaty 1990; Miller \& Roberts 1995).

We consider here the extension of these calculations to the
case of Alfv\'enic acceleration in clusters of galaxies.

If the energy injected in Alfv\'en waves is 
significantly larger than
that stored by the relativistic protons at the beginning
of the acceleration phase, then the spectrum of protons 
is expected to be considerably modified.
Under these conditions, we illustrate, in Fig.~14, the evolution 
of the spectrum of the relativistic protons. The figure clearly
shows that the spectrum flattens and develops a bump.

The prominence of this bump increases with time as the energy 
absorbed by relativistic protons also increases. Moreover, the 
bump moves toward larger momenta of the particles during the
acceleration time.

The presence of a bump in the proton spectrum may be of some
importance in the calculation of the spectrum of the high
energy secondary electrons which are expected to be produced 
during hadronic collisions.
We find that, under typical conditions in the ICM and assuming
an energetics of the Alfv\'en waves considerably larger than that 
of the initial proton population, relevant bumps are
produced at energies in excess of $200$ GeV which should be visible
in the spectrum of secondary electrons at $\gamma > 10^5$. 
We also point out that proton acceleration is strongly reduced
at some momentum $p_{max}$ (Fig.~14) which corresponds to
the momentum at which the resonance condition [Eq. (\ref{resonance})]
is satisfied for wavenumber corresponding to the maximum
injection scale of the Alfv\'en waves [Eq. (\ref{taylor})].
A maximum energy of the injected secondary electrons is expected
as well. A detailed discussion of the production of secondary 
electrons by reaccelerated protons and of their further acceleration 
will be presented in a forthcoming paper.

A relevant point to make for the process of proton acceleration 
is to quantify the fraction of the energy injected in Alfv\'en waves
which is absorbed by the relativistic protons and the time
needed for relativistic protons to react to the injection of this 
energy.
In Fig.~15, we compare the energy injected in waves with that 
converted to relativistic protons, for different values of the  
rates of energy injection in the form of Alfv\'en waves. 
After some time, it can be seen that the energy of the protons 
asymptotically converges to the energy of the waves.
It is also clear that there is a delay time between the injection
of waves and the proton acceleration, as it is expected due to the
finite acceleration time.

\subsection{The Wave-Proton Boiler}

One of the most important results of our investigation is the 
quantitative treatment of the backreaction of the accelerated 
protons on the waves and in turn on the electrons.  
Qualitatively, given typical conditions in the ICM, we can identify 
three main temporal stages of the acceleration process:

\begin{itemize}

\item[{\it 1)}] {\it Cascading stage}:
 
For a non negligible rate of energy injection in the form of
Alfv\'en waves, the cascade time is shorter than the damping time.
This remains true up to some critical wavenumber, which depends on 
energetics and spectrum of protons, where damping starts to be 
relevant.
If such a wavenumber is larger than about $10^{-2} k_{max}$, 
then enough energy is left in the form of waves at the scales 
which may resonate with relic relativistic electrons. In this
case electrons are effectively re-energized.

\item[{\it 2)}] {\it Stage of proton backreaction}:

Once the Alfv\'en waves start to accelerate electrons and protons 
to higher energies, the spectrum of protons and electrons becomes
harder and the fraction of the energy stored in non-thermal particles
starts to be large enough to make damping more severe.
As a consequence, the rate of electron acceleration is reduced.

\item[{\it 3)}] {\it End of acceleration}:

At the beginning of the acceleration phase, the bulk of protons is 
located at supra-thermal or trans-relativistic energies. It takes a 
few $10^8$yrs, however, for these protons to be energized to higher 
energies, as illustrated in Fig. 16, where we plot the acceleration 
time scale of relativistic protons (we chose a relatively steep 
injection spectrum). From Fig. 16 we see that after about $0.5-0.7$ 
Gyr the acceleration time scale has increased by about one order 
of magnitude. At this point the acceleration stage of protons and
electrons can be considered as concluded, unless the injection 
of turbulence occurs intermittently (see Sect.5.2).

After the end of the third stage, the electrons cool due to radiative 
and Coulomb losses, while the Alfv\'en acceleration is only able to prevent 
the thermalization of these particles maintaining their Lorentz factor 
around $\gamma \sim 100-1000$.

\end{itemize}

\begin{figure}
\resizebox{\hsize}{!}{\includegraphics{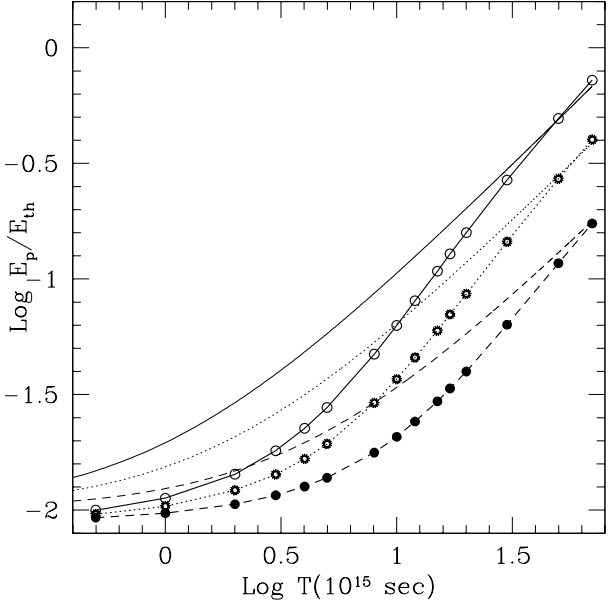}}
\caption[]{
Evolution of the energy density stored in the form of cosmic ray 
protons ($p > 0.1 \, m_p c$) as a function of time from the beginning of
the acceleration stage for three different injection rates
of Alfv\'en waves: $d(\delta B)^2 / d t = 3.3 \times 10^{-15}$
(points-solid line), $1.9 \times 10^{-15}$ (points-dotted line),
and $8.2 \times 10^{-16} (\mu {\rm G})^2$/s (points-dashed line).
For comparison the energy density injected in Alfv\'en waves
is aslo plotted. In the calculations we assume the following values of
the parameters:
$B=0.5 \mu$G, $T=10^{8}$K, $n_{th}=10^{-3}$cm$^{-3}$,
${\cal E}_e = 0.001 \times {\cal E}_{th}$, 
${\cal E}_p =0.01 \times {\cal E}_{th}$, $s=2.2$
and $z_i=1.0$.}
\end{figure}

\begin{figure}
\resizebox{\hsize}{!}{\includegraphics{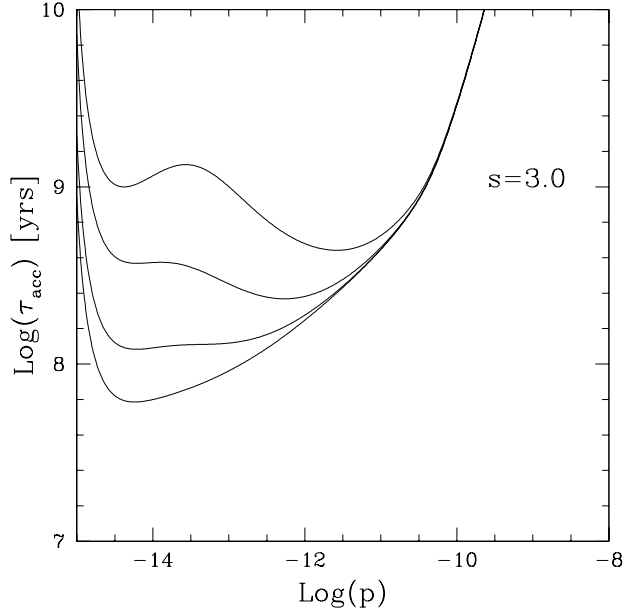}}
\caption[]{
Evolution of the proton acceleration time--scale during the acceleration 
phase as a function of $p$. From bottom to top the curves refer to 
$2\times 10^{15}$, $5 \times 10^{15}$, $10^{16}$, and $2 \times 10^{16}$ 
sec after the beginning of the acceleration stage.
The following values of the parameters have been adopted:
$d(\delta B)^2 / d t = 3.3 \times 10^{-15} (\mu {\rm G})^2$/2,
$B=0.5 \mu$G, $T=10^{8}$K, $n_{th}=10^{-3}$cm$^{-3}$,
${\cal E}_e = 0.001 \times {\cal E}_{th}$, 
${\cal E}_p =0.0025 \times {\cal E}_{th}$ and $s=3.0$.
}
\end{figure}

\section{Non-thermal emission from galaxy clusters}

\subsection{Cluster mergers and turbulence}

Major mergers are among the most energetic events in the Universe.
Cluster mergers involve a collision of at least two subclusters 
with relative velocity of $\sim 1000-2000$ km s$^{-1}$. During these
events, a gravitational energy in excess of $10^{63}$ erg is released
through the formation of shock waves. Numerical simulations show that 
cluster mergers can generate relatively strong turbulence in the ICM
(Norman \& Bryan 1999; Roettiger et al., 1999; Ricker \& Sarazin 2001).
The bulk of the turbulence is most likely injected on scales $\geq 100$ kpc
during the motion of the subclusters. Afterwards this turbulence 
eventually cascades toward smaller scales.
As discussed in Sect.3.1, when the turbulent cascade reaches scales close
to the Taylor scale, a fraction of the energy flux of the fluid turbulence 
can be transferred to MHD waves which in turn can accelerate fast particles.

For simplicity, we assume here that the bulk of the fluid turbulence 
in a given point of the cluster volume is injected at the scale $l_o$, 
for a time $\tau_i$, of the order of the time necessary for the subclump
to cross the scale $l_o$:

\begin{equation}
\tau_i ({\rm Gyr})
\sim 0.3 \xi \Big( {{l_o}\over{300}}
\Big)
\Big(
{{ v_c }\over{10^3}}
\Big)^{-1},
\label{taui}
\end{equation}
\noindent
where $v_c$ is the velocity of the subclump in the host cluster 
and $\xi$ is a parameter of the order of a few. Within these 
assumptions, the injection rate of energy in the form of fluid 
turbulence is given by :
\begin{equation}
F_f
\sim 
{{ 2.3 \times 10^{-27} }\over
{\xi}}
\Big( {{ n_{\rm th}}\over{
10^{-3}}} \Big)
\Big( {{ T }\over{10^8}}
\Big)
\Big(
{{l_o}\over{300}}
\Big)^{-1}
\Big(
{{ v_c }\over{10^3}}
\Big) 
{{ {\cal E}_t}\over{ {\cal E}_{th} }},
\label{detdt2}
\end{equation}
\noindent
where ${\cal E}_{\rm th}$ is the local energy density of the
ICM in the form of thermal gas and ${\cal E}_{\rm t}$ is that in 
the form of turbulence. The bulk of the fluid turbulence 
at the scale $l_o$ then cascades toward smaller scales producing
a spectrum of the fluid turbulence that we write as $W_f(x) 
\propto x^{-m}$ (Sect.~3.1).
Assuming a Kolmogorov phenomenology for the wave-wave diffusion
in $k-$space, the time scale for the cascade can be estimated 
from Eq. (\ref{tau-ww}) with $I_{x_o} \sim x_o^{-1} F_f$:
\begin{equation}
\tau_s ({\rm Gyr})
\sim 
0.2 
\Big(
{{l_o}\over{300}}
\Big)
\Big( {{10^8}\over{ T} }
\Big)^{1/3}
\Big(
{{10^3}\over{v_c}}
\Big)^{1/3}
\xi^{1/3}
\Big(
{{0.1}\over{ {\cal E}_t/ {\cal E}_{th}}}
\Big)^{1/3}.
\label{tau_si}
\end{equation}
\noindent
In our simple approach, this is the time delay between the merger 
event and the development of the turbulence at small scales
(and thus the production of MHD waves). In the case of Alfv\'en waves
in the framework of a Kolmogorov phenomenology, the power injected in 
waves is [Eq. (\ref{inj_alfven})]:

\begin{eqnarray}
P_A = \int I(k)dk 
\simeq \,\,\,\,\,\,\,\,\,\,\,\,\,\,\,\,\,\,\,\,\,\,\,\,\,\,\,\,\,\,
\,\,\,\,\,\,\,\,\,\,\,\,\,\,\,\,\,\,\,\,\,\,\,\,\,\,\,\,\,\,
\,\,\,\,\,\,\,\,\,\,\,\,\,\,\,\,\,\,\,\,\,\,\,\,\,\,\,\,\,\,
\nonumber\\
{{ 
1.5\times 10^{-29} }\over{
{
{\cal R}_{16}
}^{ { 1 \over 6 } }
}}
\Big( {{ v_{\rm f} }\over{400}}
\Big)^4
B_{\mu G}^{-1}
\Big({{n_{\rm th} }\over
{ 10^{-3}}} \Big)^{ {3 \over 2} }
\Big(
{{ l_o }\over{300}}
\Big)^{-1}
\label{palfven}
\end{eqnarray}
\noindent
All the quantities involved in the calculation of the power radiated 
in the form of Alfv\'en waves can be relatively well modelled. The only 
parameter which is very difficult to estimate is the value of the 
Reynolds number, ${\cal R}= l_o v_f/\nu_K$, due to the large
uncertainties in the value of the kinetic viscosity
$\nu_K = u_p \lambda_{eff}/3$ ($u_p$ is the thermal velocity and 
$\lambda_{eff}$ the effective mean free path of protons).
For transverse drift of protons in a magnetic field $B$, the mean free 
path is given by $\lambda_{eff}\sim \lambda_g^2/\lambda_c$ (e.g., 
Spitzer 1962) with $\lambda_g$ and $\lambda_c$ being the proton 
gyroradius and the mean free path due to Coulomb collisions respectively.
In this case the kinetic viscosity is given by (Fujita et al. 2003, and 
references therein):
\begin{equation}
\nu_K
= 1.3 \times 10^5 \Big(
{{ n_{th} }\over{10^{-3}}}
\Big) \Big(
{{ T}\over{10^8}}
\Big)^{-1/2}
\Big(
{{ \ln \Lambda}\over{
40}}
\Big) B_{\mu G}^{-2}.
\label{nuk}
\end{equation}

\noindent
The resulting Reynolds number is:
\begin{equation}
{\cal R}
\sim 2.8 \times 10^{26}
\Big(
{{ l_o }\over{
300}} \Big)
\Big(
{{v_f}\over{400}}
\Big)
\Big(
{ {10^{-3}}\over{ n_{th} }}
\Big)
\Big(
{{ T}\over{
10^8}}
\Big)^{ {1\over 2} }
B_{\mu G}^2
\Big(
{{40}\over{ \ln \Lambda}}
\Big),
\label{reynolds-f}
\end{equation}

which is extremely large and, most likely, should be considered
as an upper limit, due to the assumption of transverse drift of 
protons on the magnetic field lines. In general, diffusion of the
protons along the magnetic field lines can substantially
increase the value of $\lambda_{eff}$ and thus reduce ${\cal R}$.
The Reynolds number has been roughly estimated in a number of 
astrophysical situations, being $\sim 10^7$ in the solar wind 
(e.g., Grappin et al 1982), $\sim 10^{11}$ in the extragalactic 
radio jets imaged with the VLA (e.g., Henriksen, Bridle, Chan 1982),
$\sim 10^{14}$ in the solar corona (e.g., Ofman \& Aschwanden, 2002)
and $\sim 10^{16}$ for the hot phase of the local interstellar medium 
(e.g., Armstrong, Rickett, Cordes 1981). Following this last estimate, 
in our modelling we take ${\cal R} \sim 10^{16}$ but also stress that, 
due to the poor dependence of $P_A$ on ${\cal R}$, an uncertainty in  
${\cal R}$ by six orders of magnitude implies only one order of magnitude
change in $P_A$.

In our simple model for the merger, it is easy to show that the 
condition $P_A << F_f$ is satisfied, therefore the application of the
{\it Lightill} scheme in our calculations appears to be self-consistent. 

The power radiated in the form of Alfv\'en waves strongly depends on 
the velocity of the eddies in the fluid turbulence. In addition, 
assuming that the energy of the fluid turbulence is {\it a fraction 
of the local thermal energy} (i.e., ${\cal E}_t \sim \rho v_f^2$, 
and $v_f$=const) and that the largest scale of the spectrum of turbulence, 
$l_o$, does not depend on the position in the cluster volume,
we notice that the injected power in Alfv\'en waves increases with 
increasing the number density of the ICM and with decreasing the 
strength of the local magnetic field. For instance, assuming a typical 
scaling law for the magnetic field in the cluster $B\propto 
n_{\rm th}^{2/3}$, it would be $P_A \propto n_{\rm th}^{5/6}$, namely 
the power injected in the form of Alfv\'en waves is expected to be slightly
larger in the central regions of the cluster.

A second important quantity in our calculations is the largest scale of 
the spectrum of the Alfv\'en waves, $k_T \sim x_T v_f(x_T)/v_A$.
It is convenient to parametrize this quantity in terms of the
wavenumber corresponding to the minimum scale of the MHD waves, 
$k_{max} \sim \Omega_p / v_{M}$, namely:

\begin{equation}
{{ k_T}\over{k_{\rm max}}}
\simeq
{{ 
{ {\cal R}_{16}
}^{{1 \over 3} }
}\over{ 2.4 \cdot
10^{7} }}
\Big({{n_{\rm th} }\over
{ 10^{-3}}} \Big)^{ {1 \over 2} }
\Big(
{{300}\over{l_o}}
\Big)
B_{\mu G}^{-2}
\Big( {{ v_{\rm f} }\over{400}}
\Big)
\Big(
{{ T}\over{10^8}}
\Big)^{{1\over 2}}.
\label{k/kmax}
\end{equation}

Given the resonance condition [Eqs.(\ref{res_erel}) and (\ref{res_prel})], 
the presence of a maximum scale in the wave spectrum, $l_T \sim 2\pi/k_T$, 
implies a limit for the energy of the particles that can be efficiently 
accelerated:

\begin{eqnarray}
{\cal E}_{res}({\rm TeV}) \leq
24 \times 
\Big(
{{ k_{max}/k_{T} }\over
{6 \cdot 10^6 }}
\Big)
\Big(
{{ T }\over{10^8}}
\Big)^{1/2} \nonumber\\
\propto 
{\cal R}^{-1/3}
n_{th}^{-1/2}
l_o
B^2
{v_f}^{-1}.
\label{emaxres}
\end{eqnarray}

This limit provides a relatively good estimate in the case of
relativistic protons, which are basically loss-free particles, while
the maximum energy of the electrons is driven by the competition
between acceleration and loss terms.

\begin{figure*}
\resizebox{\hsize}{!}{\includegraphics{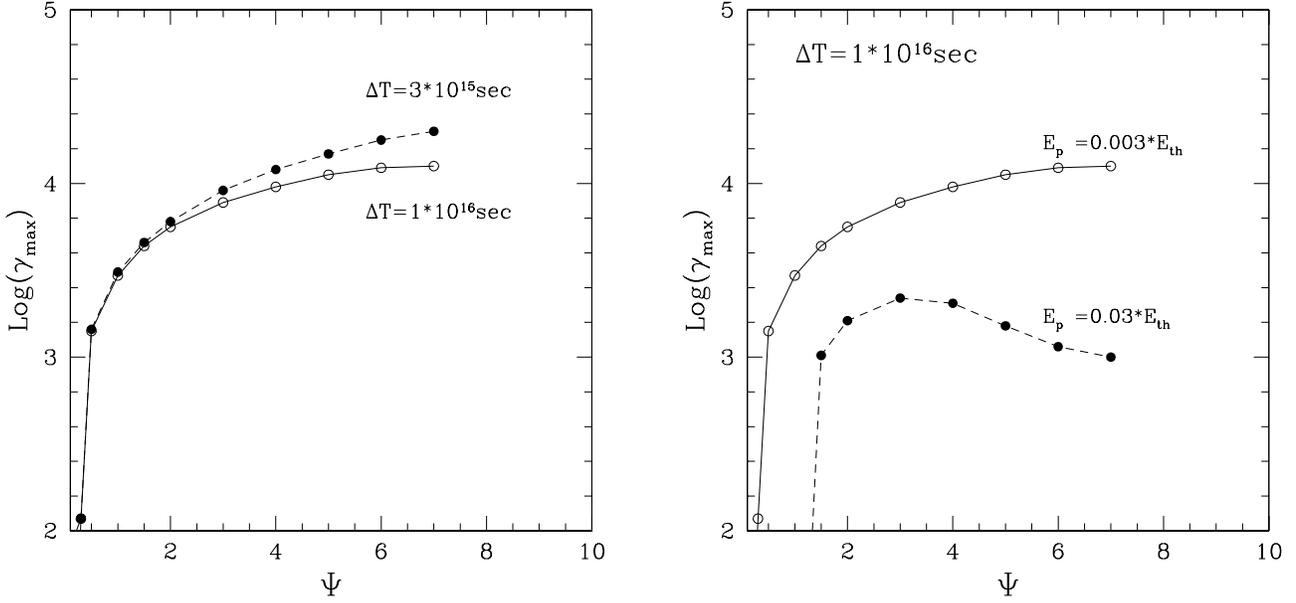}}
\caption[]{
Maximum energy attainable for relativistic electrons during the 
acceleration (only systematic terms are considered here) plotted 
as a function of $\Psi$. {\bf Left Panel}: $\gamma_{\rm max}$ 
at $3 \times 10^{15}$ (dashed line) and $10^{16}$sec
(solid line) after the beginning of the acceleration stage.
The values of the parameters are:
$B = 1 \mu$G, $n_{th}=10^{-3}$cm$^{-3}$, 
${\cal E}_e= 0.001 \times {\cal E}_{th}$,
${\cal E}_p= 0.003 \times {\cal E}_{th}$,
$s=2.2$, and $z_i=1.0$.
{\bf Right Panel}:
$\gamma_{\rm max}$ at $10^{16}$sec from the beginning of the 
acceleration stage for ${\cal E}_p= 0.003 \times {\cal E}_{th}$
(solid line) and ${\cal E}_p= 0.03 \times {\cal E}_{th}$
(dashed line). Other parameters are as in the Left Panel case.}
\end{figure*}

\subsection{Constraining the model parameters}

In this Section we derive some constraints on the
physical conditions in the ICM in order to obtain 
the reacceleration efficiency necessary 
to allow the production of the observed
non--thermal emission.

From Eq.~(\ref{palfven}) it is clear that a key parameter in 
the calculation of the efficiency of particle acceleration 
in our approach is given by :

\begin{equation}
\Psi=
\Big({{ v_f }\over{400}}
\Big)^4
\Big(
{{
{\cal R} }\over
{10^{16}}}
\Big)^{-1/6}
\Big( {{
l_o }\over
{300}}
\Big)^{-1},
\label{psi}
\end{equation}
which is very sensitive to the velocity of the eddies in the 
fluid turbulence. A constraint on this velocity can be obtained
by requiring that the energy injected in the form of fluid turbulence 
(${\cal E}_t \sim x_o W(x_o) \sim m_p n_{th} v_f^2$) does not 
exceed the thermal energy. From Eqs.~(\ref{taui}) and 
(\ref{detdt2}) we obtain:
\begin{equation}
\Big(
{{ v_f }\over{400}}
\Big)
\leq 3 \Big(
{{ T }\over{
10^8}}
\Big)^{1/2}
\Big(
{{ {\cal E}_t}\over{ {\cal E}_{th} }}
\Big)^{1/2}.
\label{limite}
\end{equation}

An additional constraint to the parameter $\Psi$ comes from
requiring that the energy stored in the form of relativistic
protons is small enough to allow for efficient electron acceleration.

As pointed out above, after an acceleration time of the order of 
0.5-0.7 Gyr, relativistic protons get a large fraction of the energy 
previously in the form of Alfv\'en waves.
We limit ourselves to cases in which 
$P_A \cdot \tau_i < 0.1 \, {\cal E}_{th}$.
From Eqs.~(\ref{taui}) and (\ref{palfven}) one has:
\begin{equation}
\Psi \leq
{{ 16}\over{\xi}} B_{\mu G}
\Big(
{{ n_{th} }\over{10^{-3}}}
\Big)^{-{1\over 2}}
\Big( {{ T}\over{10^8}}
\Big)
\Big( {{ l_o }\over{300}}
\Big)^{-1}.
\label{limitipsi}
\end{equation}

Finally, we are interested in the production of relatively 
long-living (e.g., for $> 0.3$ Gyr) non-thermal phenomena,
in order to have a chance of observing them in some clusters.
From Eq.~(\ref{taui}) we can set a limit on the parameter $\xi$:

\begin{equation}
\xi > \Big( {{v_c }\over{1000}} \Big)
\Big(
{{ l_o }\over{300}}
\Big)^{-1}.
\label{xilimit}
\end{equation}

The maximum energy, $\gamma_{max}$, of the accelerated electrons
is obtained by balancing energy losses and energy gains.

In Fig. 17a we plot $\gamma_{max}$ as a function of $\Psi$, for
different acceleration times, $\Delta T$, for a given set of
values of the parameters defining the environment of cluster
cores. In Fig. 17b we plot the same quantity for different 
values of the energy density in the form of relativistic protons, 
${\cal E}_p$. The injection spectrum of protons is taken as a 
power law with slope 2.2.

In order to obtain $\gamma_{max} >> 1000$, needed to explain the 
synchrotron emission at GHz frequency as well as the IC hard X--ray 
photons, we are forced to require that 
$\Psi \geq 1$ and ${\cal E}_p \leq 3\% \, {\cal E}_{th}$.

Such a stringent limit is the consequence of the effective 
damping of Alfv\'en waves upon the relativistic proton component,
which inhibits the acceleration of electrons.

In the periphery of the cluster, where the magnetic field
is expected to be lower, the conditions to obtain high 
energy electrons are less stringent. It remains true however
that no more than a few percent of the thermal energy of the 
cluster can be in the form of relativistic protons if we
want to interpret the observed non-thermal phenomena as the
result of radiative processes of high energy electrons
accelerated via Alfv\'en waves.

A steeper injection spectrum of protons, containing the same
energy density, makes the constraints found above even more
stringent; however this would imply a very large energy injected
in relativistic protons in the ICM.

On the other hand, at given proton number density, a steeper spectrum 
contains less relativistic particles, which would allow for more 
efficient electron acceleration. 

\begin{figure}
\resizebox{\hsize}{!}{\includegraphics{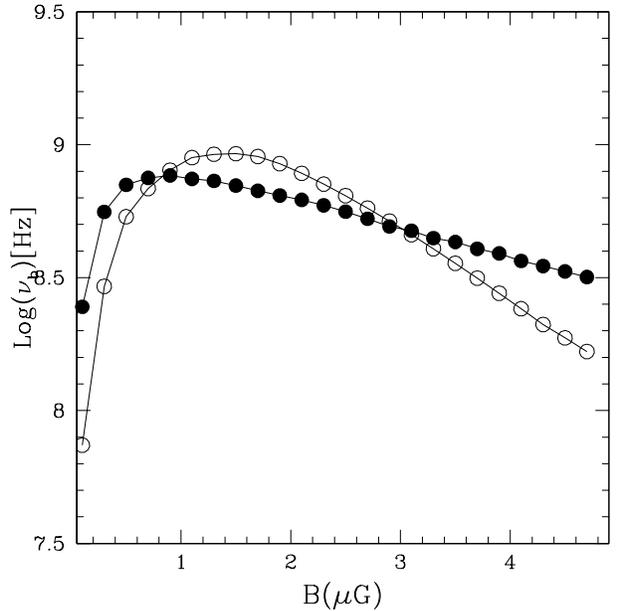}}
\caption[]{
Synchrotron cut--off frequency as a function of the
magnetic field strength in the case of 
$n_{\rm th} = 10^{-3} cm^{-3}$ (empty symbols) and
$n_{\rm th} = 2\times 10^{-4} cm^{-3}$ (filled symbols).
The results are reported after $7 \times 10^{15}$sec of
reacceleration.
In the calculations the following values of the parameters 
have been used: 
$d(\delta B)^2/d t = 3.3 \times 10^{-15} (\mu {\rm G})^2$/s, 
$T=10^8$K, ${\cal E}_e=0.001 \times {\cal E}_{th}$,
${\cal E}_p=0.01 \times {\cal E}_{th}$, $s=2.2$
and $z_i=1.0$.
}
\end{figure}

Clearly, a crucial parameter in the modeling of the non--thermal phenomena 
in galaxy clusters is the strength of the magnetic field in the ICM.
There is still debate on whether this field is of several $\mu G$ or rather
fractions of $\mu G$: on one hand, if the hard X--ray excess is interpreted 
as a result of IC emission of relativistic electrons, the required volume
averaged field is $\leq 0.2-0.4 \mu$G (e.g. Fusco--Femiano et al. 2002).
On the other hand, Faraday rotation measurements (RM) of the radiation coming 
from cluster radio sources seem to require a magnetic field strength 
of $\sim 4-8 \mu$G (e.g., Clarke et al. 2001).
A number of possibilities to reconcile these two predictions have been
proposed in the literature, based on either choosing more {\it realistic}
spectra of the radiating particles or introducing a spatial distribution 
of non-thermal particles and magnetic fields (Goldshmidt \& Rephaeli 1993;
Brunetti et al. 2001a; Petrosian 2001; Kuo et al. 2003).

It is also worth recalling that Faraday RM do not provide a direct 
measurement of the magnetic field, but rather an estimate of a 
quantity which is a (not necessarily trivial, e.g. Clarke 2002) 
convolution of the 
component of the magnetic field parallel to the line of sight, of 
the electron density and that accounts for the topology of the magnetic 
field: Newman et al. (2002) showed how the assumption of a single--scale 
magnetic field leads to an overestimate of the field strength extracted 
from RM. Other authors have further discussed the influence of the
power spectrum of the magnetic field topology in the ICM on the RM
(Ensslin \& Vogt 2003; Vogt \& Ensslin 2003; Govoni et al. 2003).

In order to illustrate the effect of these uncertainties on the 
conclusions inferred from our calculations, we evaluated the 
so-called synchrotron cut--off frequency for a cluster with 
temperature $T=10^8$ K and with a relic electron and proton energy
densities chosen as ${\cal E}_e=0.001 \times {\cal E}_{th}$, 
${\cal E}_p=0.01 \times {\cal E}_{th}$ (protons have a spectrum with 
power index $s=2.2$; see caption of Fig. 18 for additional information). 
The calculation is carried out for a dense region ($n_{\rm th} = 10^{-3} 
cm^{-3}$ [empty symbols]) and for a low density region 
($n_{\rm th} = 2\times 10^{-4} cm^{-3}$ [filled symbols]).
Given the shape of the spectrum of the accelerated electrons,
a synchrotron cut--off at $\geq 300$ MHz is required to account
for the synchrotron radiation observed in the form of radio halos.

Our conclusion, based on Fig. 18, is that in high density regions 
($n_{\rm th} \sim 10^{-3}$cm$^{-3}$) Alfv\'enic reacceleration of relic 
electrons cannot be an efficient process for $B \gg 4 \mu$G and for 
$B \ll 0.5 \mu$G. These constraints become less stringent in the case 
of low density regions.

\subsection{A simplified models for Radio Halos and Hard X--ray emission}

The best evidence for the diffuse non-thermal activity in clusters
of galaxies is provided by the extended synchrotron radio emission
observed in about $\sim 30$ massive clusters of galaxies (e.g., 
Feretti, 2002). A recent additional evidence supporting the existence
of relativistic electrons is given by the hard X--ray tails in excess 
to the thermal emission discovered by BeppoSAX and RXTE in the case 
of a few galaxy clusters (e.g., Fusco-Femiano et al. 2002).
The possibility that hard X--ray tails are due to IC scattering of the 
CMB photons is intriguing as, in this case, radio and HXR radiations 
would be emitted by roughly the same electron population and thus the 
combination of radio and HXR data would allow us to infer an estimate
of the volume-averaged magnetic field strength and of the
energy density of relativistic electrons.
Additional pieces of evidence for non--thermal phenomena are 
the so called {\it radio Relics} and the EUV excesses whose 
origin may however be not directly connected to that of radio halos 
and HXR. Therefore these phenomena will not be modelled in the present 
paper. For a recent review on these arguments the reader is referred to 
Ensslin (2002, and ref. therein) and Bowyer (2002, and ref. therein).

In this section we apply the formalism described in previous
sections in order to show that for the conditions realized in the 
ICM, Alfv\'enic reacceleration of relic electrons may generate the
observed radiation, provided the energy content in the form of 
relativistic protons is not too large. For simplicity we neglect 
here the production and reacceleration of secondary products of 
proton interactions.

\begin{figure}
\resizebox{\hsize}{!}{\includegraphics{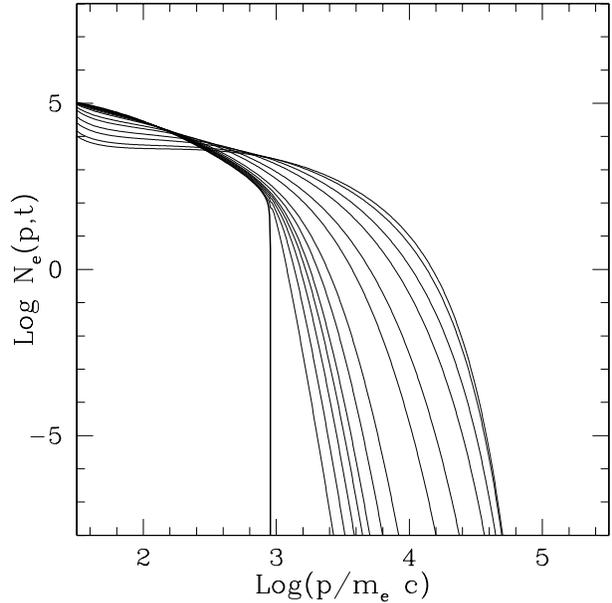}}
\caption[]{
Temporal evolution of the accelerated electron spectra after 0, 
$10^{14}$, ..., $5\times 10^{15}$, $7\times 10^{15}$,
$10^{16}$, and $1.2\times 10^{16}$sec from the beginning
of the acceleration stage.
The following values of the parameters have been used:
$d(\delta B)^2/d t = 1.6 \times 10^{-15} (\mu {\rm G})^2$/s, 
$B=0.5 \mu$G,
$T=10^8$K, $n_{th}=10^{-3}$cm$^{-3}$, 
${\cal E}_e=0.001 \times {\cal E}_{th}$,
${\cal E}_p=0.01 \times {\cal E}_{th}$, $s=2.2$
and $z_i=1.0$.}
\end{figure}

\begin{figure}
\resizebox{\hsize}{!}{\includegraphics{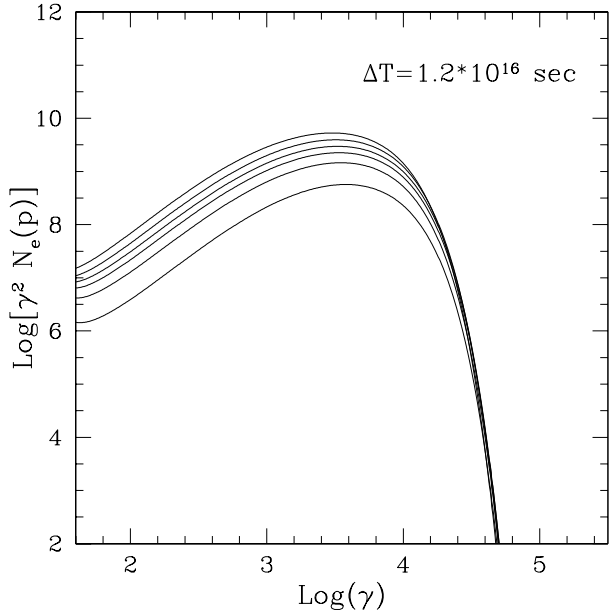}}
\caption[]{
Spectra of electrons accelerated for $1.2 \times 10^{16}$sec
at different distances from the cluster center: $r=$0.3, 0.6, 
0.9, 1.2, 1.4, 2.1$r_{\rm c}$ (from top of the diagram).
The central values assumed in the calculations are :
$n_{th}(0)=1.5 \times 10^{-3}$cm$^{-3}$, $B(0)=1.5 \mu$G, 
$d(\delta B(0))^2/d t=2.2 \times 10^{-15} (\mu {\rm G})^2$/s 
with the scaling laws given in Sect.~6.3, $r_{\rm c}=400$ kpc.
}
\end{figure}

\begin{figure*}
\resizebox{\hsize}{!}{\includegraphics{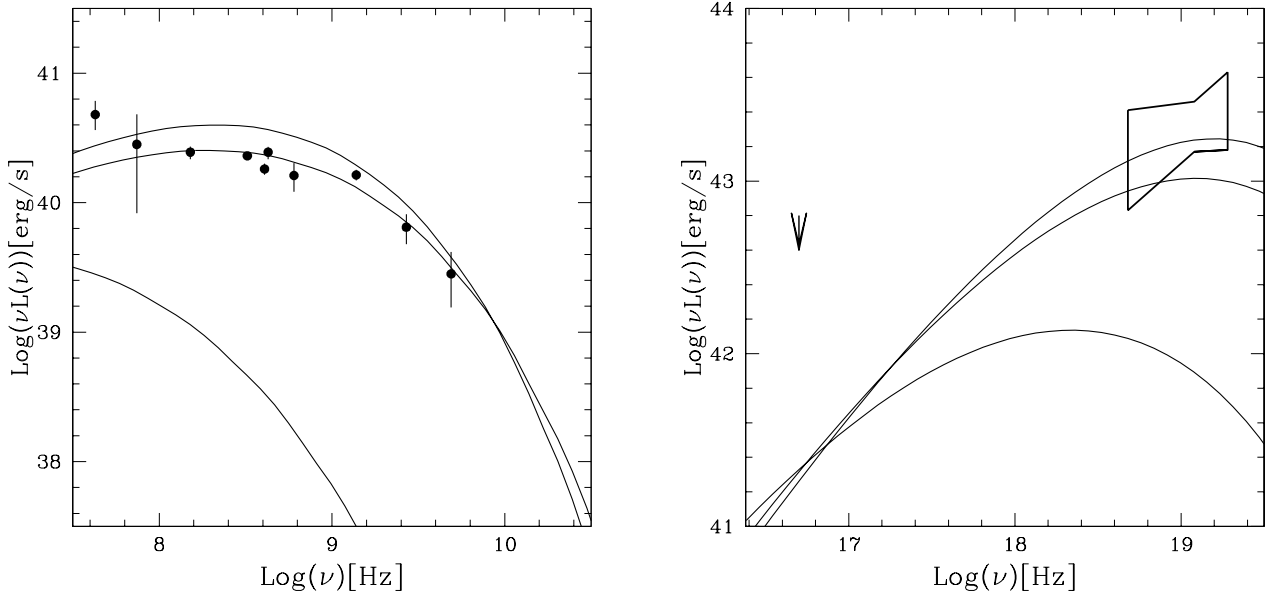}}
\caption[]{
Temporal evolution of the synchrotron ({\bf Left Panel}) and ICS 
({\bf Right Panel}) spectra from the cluster model given in Sect.~6.3
(integrating the emissivities up to $5 \times r_c$).
Spectra are shown at $5 \times 10^{15}$, $1.2 \times 10^{16}$ and 
$1.7 \times 10^{16}$sec from the beginning of the acceleration 
(from bottom to top). In the calculations ${\cal E}_e=
5\times 10^{-5} {\cal E}_{th}$ has been adopted.
Radio data are taken from Thierbach et al. (2003), EUV from 
Bowyer et al. (1999; here reported as an upper limit,
see text), XHR from Fusco--Femiano et al. (2004).}
\end{figure*}

\noindent
Our simple model for the ICM assumes a $\beta$-{\it model} 
(Cavaliere \& Fusco-Femiano, 1976) for the radial density profile of 
the thermal gas in the ICM, in the form
\begin{equation}
n_{th}(r)=n_{th}(r=0)
\Big(
1+
( {{ r }\over{r_c}} )^2
\Big)^{-3\beta /2},
\label{betamod}
\end{equation}
where $r_c$ is the core radius and we adopted $\beta = 0.8$.
The magnetic field is assumed to scale with density according 
with flux conservation:
\begin{equation}
B(r)=
B(r=0)
\Big(
{{ n_{th}(r) }\over{
n_{th}(r=0) }}
\Big)^{2/3}.
\label{b(r)}
\end{equation}
Based on the constraints given in Sect. 6.2 we adopt
$B(r=0) \sim 0.5-4 \mu$G.

Finally, we assume that the ratio between the energy density of 
the relic relativistic particles (at the beginning of the
acceleration phase) and that of the thermal plasma is constant 
with distance throughout the cluster volume:
\begin{equation}
{\cal E}_{p[e]}
={\cal E}_{th}
\eta_{p[e]},
\label{ratioenerg}
\end{equation}
where $\eta_{p[e]}$ is a free parameter ($\eta < 0.1$).

For simplicity we also assume that the maximum injection scale of 
the turbulence, the Reynolds number and the velocity of the turbulent 
eddies are independent of the location within the cluster volume.

Using the scaling relationship for the magnetic field, Eq. (\ref{b(r)}), 
and the expression for the injection power in the form of Alfv\'en waves,
Eq.~(\ref{palfven}), we obtain:

\begin{equation}
P_A(r)=P_A(r=0)
\Big(
{{ n_{th}(r) }\over{
n_{th} (r=0) }}
\Big)^{5/6}.
\label{pa(r)}
\end{equation}

The spectrum of the relativistic electrons in the core region is 
plotted in Fig. 19.
We can see that the bulk of relativistic electrons, initially
at $\gamma \sim 10^2$, can be energized up to $\gamma \sim 10^4$ 
for a relatively long time.

Eq. (\ref{pa(r)}) indicates that, in our simple approach, the power 
injected in the form of Alfv\'en waves decreases with increasing 
distance from the cluster center. Since radio halos have a considerable
size, it is needed to check that our model provides enough energy 
in the outskirts of clusters.

In Fig.~20 we plot the electron spectra at different distances 
from the cluster center (see caption). What emerges from this figure 
is that at large distances the effect of acceleration is even stronger
than in the central region and the electron spectra peak at slightly
higher energies than in the core. 
This is due to the fact that in the outskirts the 
damping rate is reduced more than the rate of injection of turbulence. 

The synchrotron emissivity roughly scales as $J_{syn} \propto 
N_e \gamma_{max}(r)^2 B^2\propto n_{th}^{7/3} \gamma_{max}(r)^2$.
Such a relatively soft dependence of the emissivity on the radius 
allows for a relatively broad synchrotron brightness profile.

More specifically, 
the synchrotron emissivity can be written as:

\begin{equation}
J_{syn}(\nu,t)
={{ \sqrt{3} e^3 B}\over{m c^2}}
\int_p \int_0^{{\pi}\over 2}
dp d\theta \sin^2\theta
N(p,t) F\Big(
{{ \nu }\over{\nu_c}}
\Big),
\label{syn}
\end{equation}

where the synchrotron {\it Kernel} is given by 
(e.g., Rybicki \& Lightman, 1979):

\begin{equation}
F\Big(
{{ \nu }\over{\nu_c}}
\Big) =
{{\nu}\over{\nu_c}}
\int_{\nu/\nu_c}^{\infty}
K_{5\over 3}(y)dy,
\label{synkernel}
\end{equation}

\noindent
and $\nu_c = (3/4\pi) p^2 e B \sin\theta /(m c)^3$.

The emissivity due to inverse Compton scattering off the CMB photons 
is given by (Blumenthal \& Gould, 1970):

\begin{eqnarray}
J_{ic}(\nu_1,t)
= \Big(
2 \pi r_o m^2 c \Big)^2
h \nu_1^2
\int_{\nu}
\int_{p}
d\nu dp 
{{ N(p,t) p^{-4} }\over
{ \exp({{ h \nu }\over{k_B T_z}}) -1 }}
\times \nonumber\\
\Big(
1+ 
2 \ln\Big({{ \nu_1 m^2 c^2}\over{ 4 p^2 \nu}}\Big)+
{{ 4 p^2 \nu}\over
{m^2 c^2 \nu_1}}
-
{{ m^2 c^2 \nu_1}\over{
2 p^2 \nu}}
\Big),
\label{ic}
\end{eqnarray}

\noindent
where $T_z=2.73 (1+z)$ is the temperature of the CMB
photons.

The corresponding synchrotron and IC spectra are plotted in Fig. 21
at different times, for a central magnetic field $B(r=0) \simeq 1 
\mu G$.
The spectra are compared with that observed for the radio halo
in the Coma cluster : an initial energy density in the 
relic relativistic electrons of the order of 
$5 \times 10^{-5} {\cal E}_{th}$ is required to account for the data.
In Fig. 21 we also report the luminosity of the EUV excess in the
Coma cluster as an upper limit since it is commonly accepted that
the origin of the EUV excess is not directly related to the same
electron population responsible for the radio and possibly for the
HXR emission (Bowyer \& Bergh\"ofer 1998; Ensslin et al. 1999;
Atoyan \& V\"olk 2000; Brunetti et al. 2001b; Tsay et al. 2002).
Here we stress that Fig. 21 does not show the best fit
to the data but just a comparison between data and time evolution
of the emitted spectra resulting from the very simple scaling 
of the parameters described above. 
On the other hand, it should also be stressed that the time evolution
of the synchrotron and IC spectra reported in Fig. 21 is generated 
via the first fully self-consistent calculation of particle 
acceleration in galaxy clusters. Thus provided that the energy of
relativistic protons in galaxy clusters is not larger than a few percent 
of the thermal energy, Fig. 21 proves, the possibility
to obtain the observed magnitude of the non--thermal emission in
these objects via Alfv\'enic acceleration.

In passing we note that the integrated radio spectrum at $\sim$ GHz
frequencies is steeper than the IC spectrum in the hard X--ray
band independently of the time slice. This seems in nice agreement
with some temptative evidences published in some recent literature
(Fusco-Femiano et al., 2000; 2002).

\section{Conclusions}

The origin of the extended non-thermal emission in galaxy clusters
is still a subject of active investigation. Despite the numerous
models put forward to explain the observed features of this 
diffuse emission, only some of them appear to be able to describe
the observations in detail. One of the approaches that 
has definitely produced the most promising results consists in
the continuous reacceleration of relic relativistic electrons
leftover of the past activity occurred within the ICM. The 
reacceleration is likely to occur through resonant interaction
of electrons and MHD waves. 

In this paper we presented a full account of the time-dependent 
injection of fluid turbulence, its cascade to smaller scales, the 
radiation of Alfv\'en waves through the {\it Lighthill} mechanism, the 
resonant interaction of these waves with electrons and protons 
(namely their acceleration) and the backreaction of the accelerated
particles on the waves. 

The solution of the coupled evolution equations for the electrons,
protons and waves revealed several new interesting effects resulting
from the interaction among all these components:

{\it i)} Alfv\'en waves are radiated by the fluid turbulence through
the {\it Lighthill} mechanism. This allows us to establish a direct
connection between the fluid turbulence likely to be excited during
cluster mergers, and the MHD turbulence that may resonate with particles
in the ICM.

{\it ii)} Previous calculations looked for self-similar solutions 
for the spectra of electrons and MHD waves in the form of power laws.
The solutions obtained in the present paper show that in Nature 
these self-similar solutions are not necessarily achieved. In 
general the system evolves toward complex spectra of electrons and  
MHD waves with a bump in the electron spectrum, that moves in 
time toward an increasingly large particle momentum.

{\it iii)} We limit ourselves here with the case of Alfv\'en waves, 
which couple efficiently with both thermal and relativistic protons, 
but only with relativistic electrons. The main damping of Alfv\'en
waves occurs on relativistic protons, if there are enough of them.
The spectrum of the waves is cutoff at small scales due to the damping
of these waves. The damping moves energy from the waves to the 
particles, determining their acceleration/heating.

{\it iv)} The importance of the presence of the relativistic protons 
for the acceleration of electrons is one of the most relevant new
results of this work. A large fraction of the thermal energy in the
form of relativistic protons enhances the damping rates of Alfv\'en waves,
suppressing the possibility of resonant interaction of these waves 
with electrons. Since electrons are the particles that radiate the most,
a too large fraction of relativistic protons suppresses non-thermal 
phenomena directly related to electron reacceleration via Alfv\'en resonance. 
It is worth reminding that in principle the secondary electrons and
positrons resulting from hadronic interactions may also be re-energized
by waves. We neglect this effect here.
Our results show that no more than a few percent of
the thermal energy density of the cluster can be in the form of 
relativistic protons if we want to interpret the diffuse radio 
and hard X-ray emissions as the result of synchrotron and ICS
radiation of relic electrons reaccelerated through Alfv\'en waves. 
This appears as a stringent constraint on the combination
of proton number and spectrum since the accumulation of 
large number of protons in the ICM is predicted by both analytical 
calculations (Berezinsky, Blasi \& Ptuskin 1997) and numerical 
simulations (Ryu et al. 2003). The amount of energy piled up
in the form of protons in the ICM depends strongly upon the 
history of cosmic ray injection: the ICM is polluted with cosmic 
ray protons both due to single sources such as galaxies, radio 
galaxies and active galaxies, but also due to diffuse acceleration 
events, such as mergers of clusters of galaxies and accretion of 
cosmological gas onto a gravitational potential well of a cluster 
which has already been formed. The spectrum of the cosmic rays
diffusively trapped in the ICM depends upon their origin. It is
expected that mergers may contribute a major fraction of these cosmic 
rays, but the spectrum has been calculated to be relatively steep
because of the weakness of the merger related shocks. Accretion 
on the other hand, contributes spectra which are as flat as $E^{-2}$. 
Both the energy and spectrum of the cosmic rays stored in the ICM 
affect the temporal evolution of Alfv\'en waves and their ability 
to reaccelerate relic electrons. 
If future observations will unveil the presence of a population of 
relativistic protons in the ICM with $> 5-10$\% of the thermal energy, 
then Alfv\'enic reacceleration of relativistic electrons in galaxy clusters
will be discarded as a possible explanation of non-thermal phenomena
in the ICM. On the other hand, similar reacceleration phenomena can be
driven by MHD waves other than Alfv\'en waves (e.g., MS, LH), for which 
the energy transfer does not occur preferentially toward protons. 
In this case, the bounds presented here on the allowed energy 
density in the form of relativistic protons in the ICM could be substantially
relaxed.

{\it v)} 
By assuming that protons have a relatively flat spectrum ($s \sim 2.2$)
and that they contain up to a few percent of the thermal energy,  
we used a simple but phenomenologically well motivated
model for the density and magnetic field in a cluster of galaxies
in order to calculate the expected non-thermal radio and hard X-ray
activity of the cluster as a function of time. 
For the first time we performed a fully self consistent
calculation and showed that the reacceleration of relic electrons 
through resonant interaction with Alfv\'en waves can explain very well 
the observed phenomena, including the extended diffuse appearance of this 
emission. 
In passing, without considering the case of the HXR emission, 
we also showed that a magnetic field strength in the range $0.5-4 \mu$G 
in the cluster cores allow electron acceleration efficient enough to 
produce GHz synchrotron emission. These conditions are less stringent in
the outermost regions of the clusters.

{\it vi)}
In our calculations we adopted a constant injection rate of 
turbulence during particle acceleration.
In this case, even assuming 
${\cal E}_p < 0.1 \times {\cal E}_{th}$, we find
that after a few $10^8$yrs of 
acceleration the 
backreaction of protons on MHD waves 
can suppress the acceleration of energetic electrons :
this provides a limit on 
the duration of the non-thermal phenomena in galaxy clusters.
It is possible to extend the duration of non-thermal activity
assuming that injection of turbulence occurs in relatively
short bursts of duration comparable with the life--time
of electrons.
On the other hand, independently on the assumptions,
our results show that if particle
acceleration is mainly due to Alfv\'en waves,
then the existence of radio halos and HXR tails
in massive clusters should be limited to achieve 
periods of $< 10$\% of the Hubble time.

{\it vii)} 
The temporal duration of the process of turbulent cascade, and the 
acceleration time scale of electrons are estimated to be about one 
order of magnitude shorter than the dynamical time-scale of a merger 
event. This implies that a temporal correlation is still expected 
between merging processes and the rise of the non-thermal phenomena 
in galaxy clusters.

\section{Acknowledgements}

We thank L. Feretti and G. Setti for useful discussions and an 
anonymous referee for useful comments on the manuscript. 
G.B. and R.C. acknowledge partial support from CNR grant 
CNRG00CF0A. The research activity of P.B. and S.G. is partially 
funded through grants Cofin-2002 and Cofin-2003.

\end{document}